\documentclass[a4paper, fleqn, usenatbib]{mnras}
\usepackage{graphicx}	
\usepackage{amssymb}
\usepackage{natbib}
\usepackage{float}
\usepackage{amsmath}  
\usepackage{comment}
\usepackage{booktabs}
\usepackage{pdfpages}
\usepackage{epstopdf}

\newcommand{\degree}{\ensuremath{^\circ}}

\title{Temporal correlation between the optical and $\gamma$-ray flux
variations in the blazar 3C 454.3}

\author[Bhoomika et al.]{Bhoomika$^{1}$\thanks{E-mail: bhoomika@iiap.res.in},
C. S. Stalin$^{1}$,
S. Sahayanathan$^{2}$, 
Suvendu Rakshit$^{3,4}$, Amit Kumar Mandal$^{1,5}$
\newauthor 
\\\\
$^{1}$Indian Institute of Astrophysics, Block II, Koramangala, Bangalore 560034, India\\
$^{2}$Astrophysical Sciences Division, Bhabha Atomic Research Centre, Mumbai, India\\
$^{3}$ Finnish Centre for Astronomy with ESO (FINCA), University of Turku, Quantum, Vesilinnantie 5, 20014 University of Turku, Finland\\
$^{4}$Astrophysical Sciences Division, Bhabha Atomic Research Centre, Mumbai, India\\
$^{5}$Department of Physics, CHRIST (Deemed to be University), Hosur Road, Bangalore 560029, India\\
}

\date{}

\pubyear{2018}

\begin{document}
\label{firstpage}
\pagerange{\pageref{firstpage}--\pageref{lastpage}}
\maketitle


\begin{abstract}

Blazars show optical and $\gamma$-ray flux variations that are generally
correlated, although there are exceptions. Here we present anomalous
behaviour seen in the blazar 3C 454.3 based on an analysis of
quasi-simultaneous data at optical, UV, X-ray and $\gamma$-ray energies,
spanning about 9 years from August 2008 to February 2017. We have
identified four time intervals (epochs), A, B, D and E, when the
source showed large-amplitude optical flares. In epochs A and B the
optical and $\gamma$-ray flares are correlated, while in D and E
corresponding flares in $\gamma$-rays are weak or absent.  In epoch B the
degree of optical polarization strongly correlates with changes in
optical flux during a short-duration optical flare superimposed on one
of long duration. In epoch E the optical flux and degree of
polarization are anti-correlated during both the rising and declining
phases of the optical flare.  We carried out broad-band spectral
energy distribution (SED) modeling of the source for the flaring
epochs A,B, D and E, and a quiescent epoch, C. Our SED modeling
indicates that optical flares with absent or weak corresponding $\gamma$-ray
flares in epochs D and E could arise from changes in a combination of
parameters, such as the bulk Lorentz factor, magnetic field and
electron energy density, or be due to changes in the location of the
$\gamma$-ray emitting regions.

\end{abstract}




\section{Introduction}
Blazars are a peculiar class of active galactic nuclei (AGN) that have their relativistic jets pointed close to the line of sight to the observer with angle $\le$ 10$^{\degree}$ \citep{1993ARA&A..31..473A,1995PASP..107..803U}. They are classified as flat spectrum radio quasars (FSRQs) and BL Lacerate (BL Lac) objects based on the strength of the emission lines in their optical/infrared (IR) 
spectrum. Both classes of objects emit radiation over the entire accessible electromagnetic spectrum from low energy radio to high energy $\gamma$-rays. As blazars are aligned close to the observer, the emission is highly Doppler boosted causing them to appear as bright sources in the extra-galactic sky. They dominate the extragalactic $\gamma$-ray sky first hinted by the Energetic Gammma-ray Experiment Telescope (EGRET) observations on board the {\it Compton Gamma-Ray Observatory} 
(CGRO; \citealt{1999ApJS..123...79H}) and now made apparent by the Large Area Telescope (LAT) onboard the {\it Fermi} Gamma-ray space telescope \citep{2009ApJ...697.1071A}. The broad band spectra of blazars is dominated by emission from the jet with weak or absent emission lines from the broad line region (BLR). One of the defining characteristics of blazars is that they show flux variations \citep{1995ARA&A..33..163W} over a wide range of wavelengths on timescales ranging from months to days and minutes. In addition to flux variations they also show large optical and radio polarization as well as optical polarization variability. In the radio band they have flat spectra with the radio spectral index ($\alpha_r$) $<$ 0.5 ($S_{\nu} \propto \nu^{-\alpha_r})$. The broad band spectral energy distribution (SED)  of blazars is characterized  by a two hump structure, one peaking at low energies in the optical/IR/X-ray region and the other one peaking at high energies in the X-ray/MeV region \citep{1998MNRAS.299..433F,2016ApJS..224...26M}. In the one-zone leptonic emission models, the low energy hump is due to synchroton emission processes and the high energy hump is due to inverse Compton
(IC) emission processes \citep{2010ApJ...716...30A}. The seed photons for the IC process can be either internal to the jet (synchroton self Compton or SSC; 
\citealt{1981ApJ...243..700K,1985ApJ...298..114M,1989ApJ...340..181G} ) or external to the jet (external Compton or EC; \citealt{1987ApJ...322..650B}). In the case of EC, the seed photons can be from the disk \citep{1993ApJ...416..458D,1997A&A...324..395B}, the BLR \citep{1996MNRAS.280...67G,1994ApJ...421..153S} and the torus\citep{2000ApJ...545..107B,2008MNRAS.387.1669G}.  Though leptonic models are found to fit the observed SED of majority of blazars, for some blazars, their SEDs are also well fit by either hadronic \citep{2003APh....18..593M,2013ApJ...768...54B} or lepto-hadronic models \citep{2016ApJ...826...54D,2016ApJ...817...61P}. In the hadronic scenario, the $\gamma$-ray emission is due to synchroton radiation from extremely relativisitic protons \citep{2003APh....18..593M} or the cascade process resulting from proton-proton or proton-photon interactions 
\citep{1993A&A...269...67M}. Even during different brightness/flaring states of a source, a single emission model is not able to fit the broad band SED at all times. For example in the source 3C 279, while the flare during March - April 2014 is well fit by leptonic model \citep{2015ApJ...803...15P}, the flare in 2013 December with a hard $\gamma$-ray spectrum is well described by lepto-hadronic processes \citep{2016ApJ...817...61P}. Thus, the recent availability of multiwavelength data coupled with studies of sources at different active states indicate that we still do not have a clear understanding of the physical processes happening close to the central regions of blazars.

An alternative to the SED modeling approach to constrain the emission models in blazars is though multiband flux monitoring observations. In the leptonic scenario of emission from the jets of blazars \citep{2007Ap&SS.309...95B}, a close correlation between the optical and 
$\gamma$-ray flux variations is expected. However, in the hadronic scenario of emission from blazars \citep{2001APh....15..121M}, optical and $\gamma$-ray flux variations may not be correlated. Thus, optical and $\gamma$-ray flux variability observations could constrain the leptonic v/s hadronic emission model of blazar jets. Recent observations made with the Fermi Gamma-ray space telescope \citep{2009ApJ...697.1071A} coupled with observations in the optical and infrared wavelengths indicate that in majority of the blazars studied for flux variations, $\gamma$-ray flares are closely associated with flares detected at the optical wavelengths with or without lag \citep{2009ApJ...697L..81B,2012ApJ...749..191C, 2014ApJ...783...83L,2015MNRAS.450.2677C}. However the availability of good time resolution of optical and $\gamma$-ray lightcurves has led to the identification of isolated flaring events in optical and $\gamma$-rays termed as "orphan" flares. Both orphan $\gamma$-ray flares (prominent flare in GeV band $\gamma$-rays with no corresponding flare in the optical band) and orphan optical flares (flaring event in the optical band with no counterpart in the $\gamma$-ray  band) are now known in blazars. As of today, optical flares with no corresponding $\gamma$-ray flares, are known in PKS 0208-5122 \citep{2013ApJ...763L..11C} and S4 1849+67 \citep{2014ApJ...797..137C} and $\gamma$-ray flares with no corresponding optical flares, are known in PKS 2142-75 \citep{2013ApJ...779..174D}, PKS 1510-089  \citep{2015ApJ...804..111M} PKS 0454-234 \citep{2014ApJ...797..137C} and 3C 454.3 \citep{2011ApJ...736L..38V}. We are carrying out a systematic analysis of the multiwavelength variability characteristics of a sample of blazars to (i) identify anomalous flux variability behavior in blazars and (ii) constrain the physical processes 
happening in the central regions of blazars using flux variability and broadband SED modeling. Here, we present our results on the brightest source in our sample 3C 454.3. 

3C 454.3 is a FSRQ at a redshift $z$ = 0.859. It was detected first as a bright and variable $\gamma$-ray source by EGRET onboard CGRO \citep{1993ApJ...407L..41H}. It has been studied extensively utilizing data over a large range of wavelengths that include, optical, X-ray and $\gamma$-ray energies \citep{2009ApJ...697L..81B,2012ApJ...756...13B,2010ApJ...721.1383A,2017MNRAS.464.2046K}. 3C 454.3 was found in a highly active state in the $\gamma$-ray band by AGILE \citep{2010ApJ...712..405V,2009ApJ...690.1018V} in 2007. In 2010 November the highest flare was detected at E>100 MeV with the LAT instrument, having a flux value of about 
6.6 $\times$ $10^{-5}$ph $cm^{-2}$ $s^{-1}$\citep{2011ApJ...733L..26A}. 
According to \cite{2017MNRAS.470.3283S} X-ray and $\gamma$-ray emission from  3C 454.3 cannot be explained by single emission mechanism and to study 
the high energy observations one needs to consider both EC and 
SSC emission processes.  We present here our results on the  multiwavelength 
analysis carried out on 
the blazar 3C 454.3 using data that spans about 9 years from 2008 August to 
2017 February with the prime motivation to find  possible correlation between 
optical and $\gamma$-ray flux variations and subsequently constrain the
emission processes in its central region.  
In section  \ref{sec:data}, we present the data used in this work. The analysis 
are described in Section \ref{sec:analysis}, followed by the results
 and discussion in Section \ref{sec:results}. The results are summarized in 
Section \ref{sec:summary}.

\begin{figure*}
\hspace*{-3cm}
\includegraphics[width=1.3\textwidth]{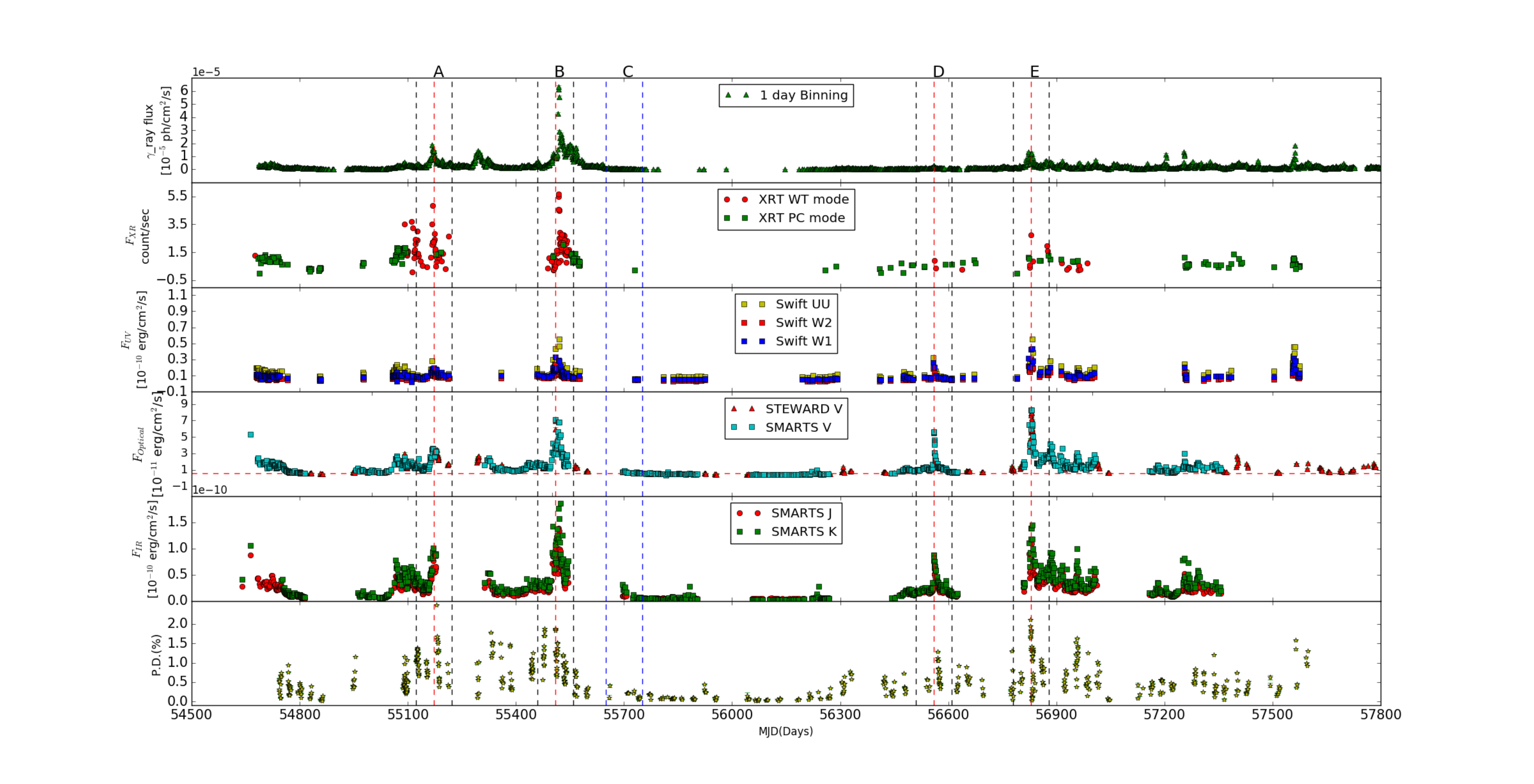}
\caption{Multiwavelength Lightcurve of the source 3C 454.3. From the top, the 
first panel shows the 
1 day binned $\gamma$-ray lightcurve for the time range 
MJD 54500-57800; the  second panel shows the SWIFT-XRT lightcurve in both 
PC(photon counting) and WT(Window timing) modes, the third panel shows the 
Swift UVOT lightcurves in W1, W2 and UU bands; in the fourth panel optical 
light curve in V-band is given; the fifth panel shows the IR light curves in J 
and K bands and in the bottom panel variation of the degree of polarization 
is presented.The red vertical lines correspond to the peaks of the optical flares and the two black vertical lines denote a width of 50 days each on either side of the peak of the flare. The two vertical blue lines have a width of 100 days and correspond to the quiescent period.}
\label{Fig1}
\end{figure*}

\section{Multi-wavelength data and reduction} \label{sec:data}

We used all the publicly available multi-wavelength flux monitoring data in 
$\gamma$-rays, X-rays, optical, UV and IR bands that span about 9 years 
covering the period 2008 August to 2017 February. We also used optical 
polarimetric data that were available during the above period. 

\subsection{$\gamma$-ray data}
For $\gamma$-rays, we used the data from the LAT instrument on-board {\it Fermi.} The {\it Fermi}-LAT is a pair-conversion telescope sensitive to $\gamma$-ray 
energies from 20 MeV to more than 300 GeV \citep{2009ApJ...697.1071A}.  {\it Fermi} normally operates in scanning mode and covers the entire sky once 
every$\sim$3 hr. We used all data for 3C 454.3 collected for the period Aug 2008 to Feb 2017 (MJD: 54500 -57800; $\sim$110 months) and analysed using the Fermi Science Tool version v10r0p5 with appropriate selections and cuts recommended for the scientific analysis of PASS8 data \footnote{ http://fermi.gsfc.nasa.gov/ssc/data/analysis/documentation/}. 
The photon-like events categorized as 'evclass=128, evtype=3' with energies 0.1$\leqslant$E$\leqslant$300 GeV $\gamma$-rays within a  circular region of interest (ROI) of 15$^\circ$ centered on the source and with zenith angle 90$^\circ$ were extracted. The appropriate good time intervals were then generated by using the recommended criteria "(DATA$\_$QUAL > 0)\&\&(LAT$\_$CONFIG==1)". The likely effects of cuts and selections, as well as the presence of other sources in the ROI, were incorporated by generating exposure map on the ROI and an additional 
annulus of 15$^\circ$ around it with the third LAT 
catalogue (3FGL - gll$\_$psc$\_$v16.fit; 
\citealt{2015ApJS..218...23A}). We used the latest isotropic model, "iso$\_$P8R2$\_$SOURCE$\_$V6$\_$v06" and the Galactic diffuse emission model "gll$\_$iem$\_$v06" . To evaluate the significance of detection, we used the maximum-likelihood (ML) ratio test defined as  TS = 2$\bigtriangleup$ log(L), where L is the 
likelihood function between models with and without a $\gamma$-ray point source at the position of  3C 454.3 \citep{2015AJ....149...41P}. We considered the source as detected if TS $>$ 9, which corresponds to a 3$\sigma$ detection 
\citep{1996ApJ...461..396M}. We generated the source light curve with a time binning of 1 day. For bins with TS $<$ 9, the source was considered undetected. We arrived at a $\gamma$-ray light curve containing 2394 confirmed measurements. All the errors associated with {\it Fermi}-LAT points are the 1$\sigma$ statistical uncertainties.

\begin{figure*}
\vspace*{-2.0cm}
\hspace*{-2.0cm}\includegraphics[width=1.2\textwidth]{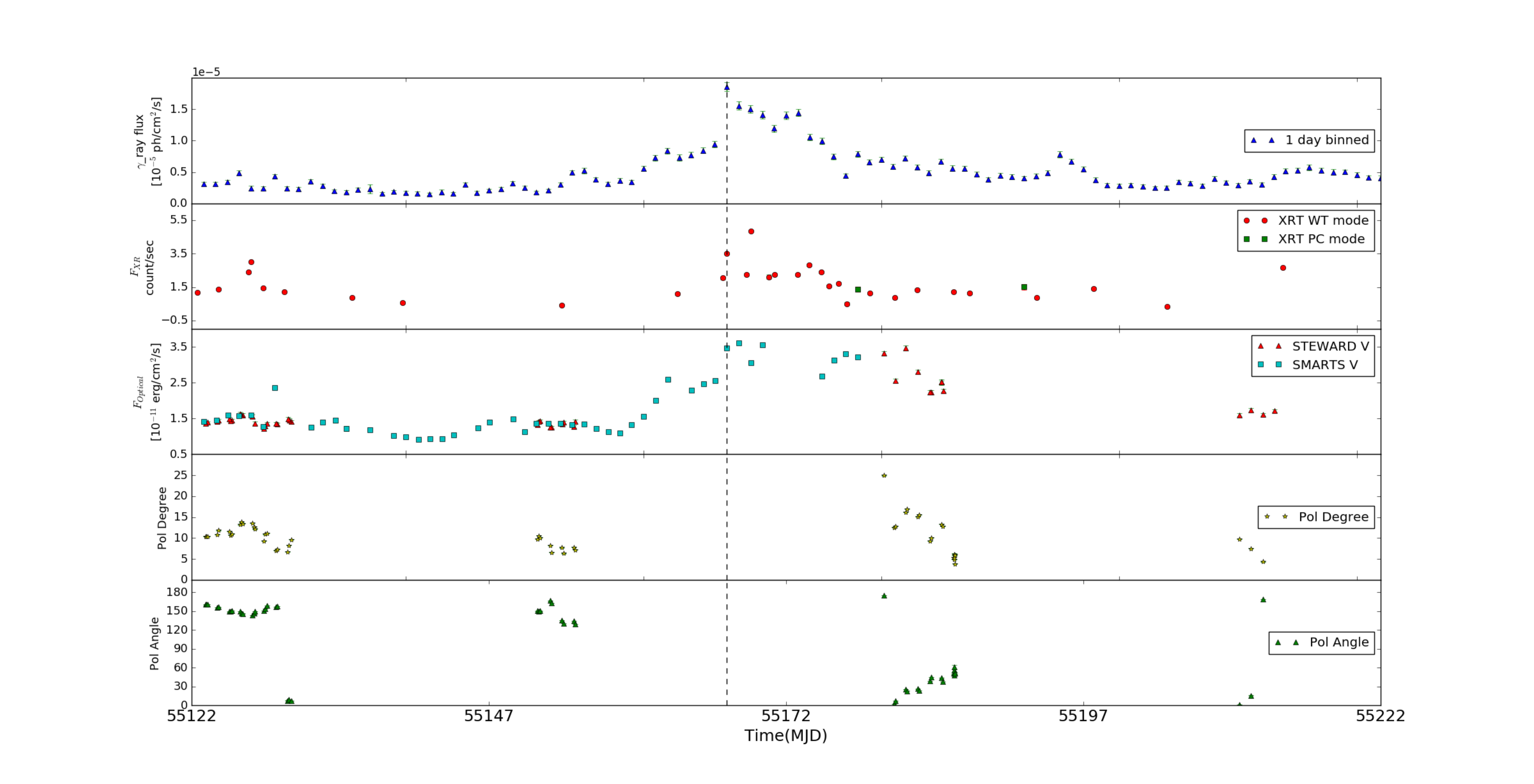}
\caption{Multi-wavelength light curves covering a period of 100 days during
epoch A. Here, from the top the first panel shows the $\gamma$-ray variations,
the second and third panels show the variations in X-ray and optical bands and
the bottom two panels show the variations in degree of optical polarization 
and polarization position angle.}
\label{Fig2}
\end{figure*}

\subsection{X-ray data}
For X-rays covering the energy range of 0.3 $-$ 10 keV, we used the data from the X-ray Telescope (XRT; \citealt{2005SSRv..120..165B}) onboard the {\it Swift} satellite \citep{2004ApJ...611.1005G} taken from the archives at HEASARC
\footnote{https://heasarc.gsfc.nasa.gov/docs/archive.html}. The data collected during the period 2008 August - 2017 February were analyzed with default 
parameter settings following the procedures given by the instrument team. For light curve analysis, data collected using both  window timing (WT) and photon counting (PC) modes were used, however, for spectral analysis only data collected from PC mode were used. The collected  XRT data were processed with the xrtpipeline task using the latest CALDB files available with version HEASOFT-6.21.We used the standard grade selection 0-12. The calibrated and cleaned events files were summed to generate energy spectra. The source spectra were extracted from a circular region of radii 50", whereas, the background spectra were selected from the region of radii 55". We combined the exposure map using the tool XIMAGE and ancillary response files created with xrtmkarf. We used an absorbed simple power law model with the Galactic neutral hydrogen column density of $N_{H}$=6.5$\times$ $10^{20}$  $cm^{-2}$ from \cite{2005A&A...440..775K} to perform the fitting within XSPEC \citep{1996ASPC..101...17A}. We obtained 128 and 191  X-ray flux  measurements
at 0.3 - 10 keV in WT and PC mode respectively.

\begin{figure}
\hspace*{-0.2cm}\includegraphics[scale=0.7]{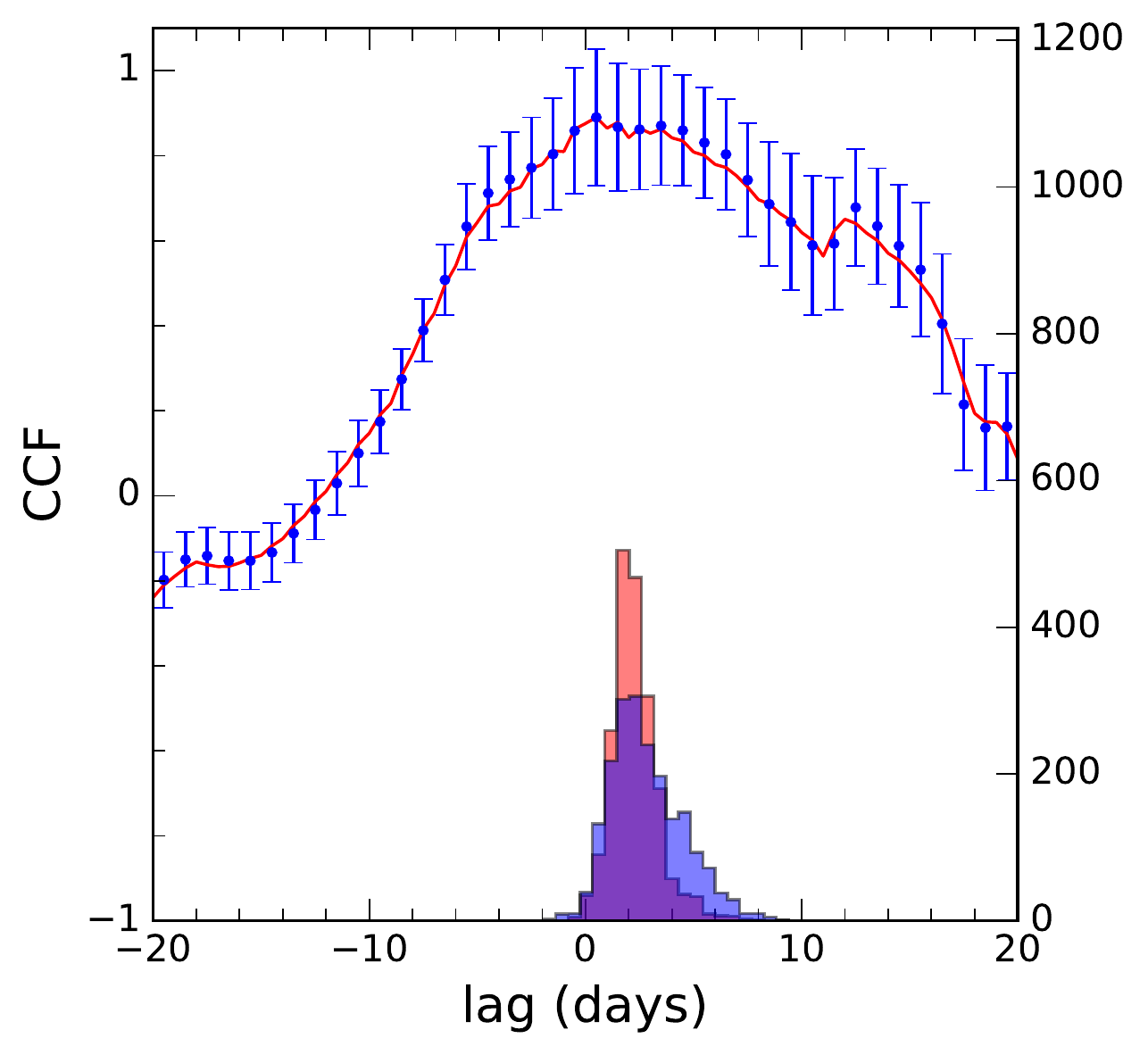}
\caption{Cross-correlation analysis between $\gamma$-ray and  optical 
flux variations during epoch A. The solid line is for ICCF and the filled
circles refer to DCF. The histograms in blue and orange show the distribution 
of cross-correlation centroids for ICCF and DCF respectively.}
\label{Fig3}
\end{figure}

\subsection{UV-Optical-NIR Data}
For UV and optical we used the data from the Swift-UV-Optical telescope (UVOT). We integrated Swift-UVOT data using the task {\tt uvotimsum}. Source counts were extracted from a 5" circular region centered at the target and background region was extracted from a larger area of 10" from a nearby source free region. The magnitude of 3C 454.3 was extracted using the task {\tt uvotsource}. The magnitudes were not corrected for Galactic reddening. The estimated magnitudes were then converted to flux units using the  zero points given in \cite{2011AIPC.1358..373B}.  In addition to UVOT, optical data were also taken from SMARTS
\footnote{http://www.astro.yale.edu/smarts/glast/home.php} (Small and Moderate
Aperture Research Telescope System) as well as the 
Steward Observatory \footnote{http://james.as.arizona.edu/$\sim$psmith/Fermi}.
Similarly near-infrared (NIR) data in J and K bands were taken from observations
carried out using the ANDICAM instrument on the SMARTS 1.3 m telescope as part of a blazar monitoring campaign, supporting the {\it Fermi} multiwavelength AGN science. The details of the instrument and the data reduction procedures can be found in \cite{2012ApJ...756...13B}.


\begin{figure*}
\vspace*{0.2cm}
\hspace*{-2.0cm}\includegraphics[width=1.2\textwidth]{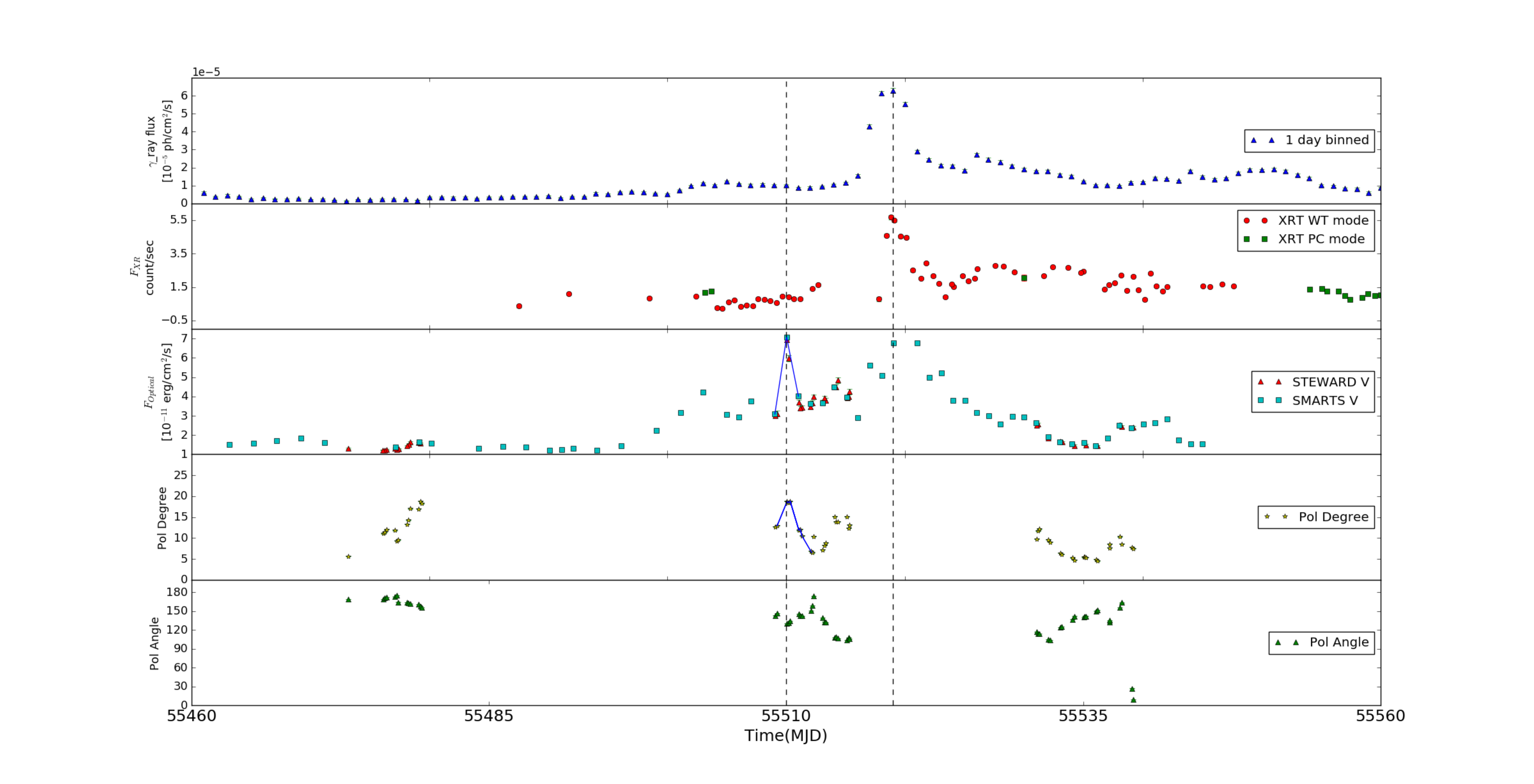}
\caption{Multi-wavelength light curves for a duration of 100 days during
epoch B. The panels have the meanings as that of Fig. \ref{Fig2}.}
\label{Fig4}
\end{figure*}

\subsection{Optical polarization data}
Optical polarization data in the V-band were taken from  Steward Observatory of the University of Arizona. Details of the data and its reduction based on spectropolarimetric observations can be found in \cite{2009arXiv0912.3621S}. The polarization data available from Steward observatory and covering the period 2008 
August to 2017 February consisted of 644 measurements. The optical polarization data along with flux measurements in other wavebands are shown in Fig. \ref{Fig1}.

\begin{figure}
\hspace*{-0.2cm}\includegraphics[scale=0.7]{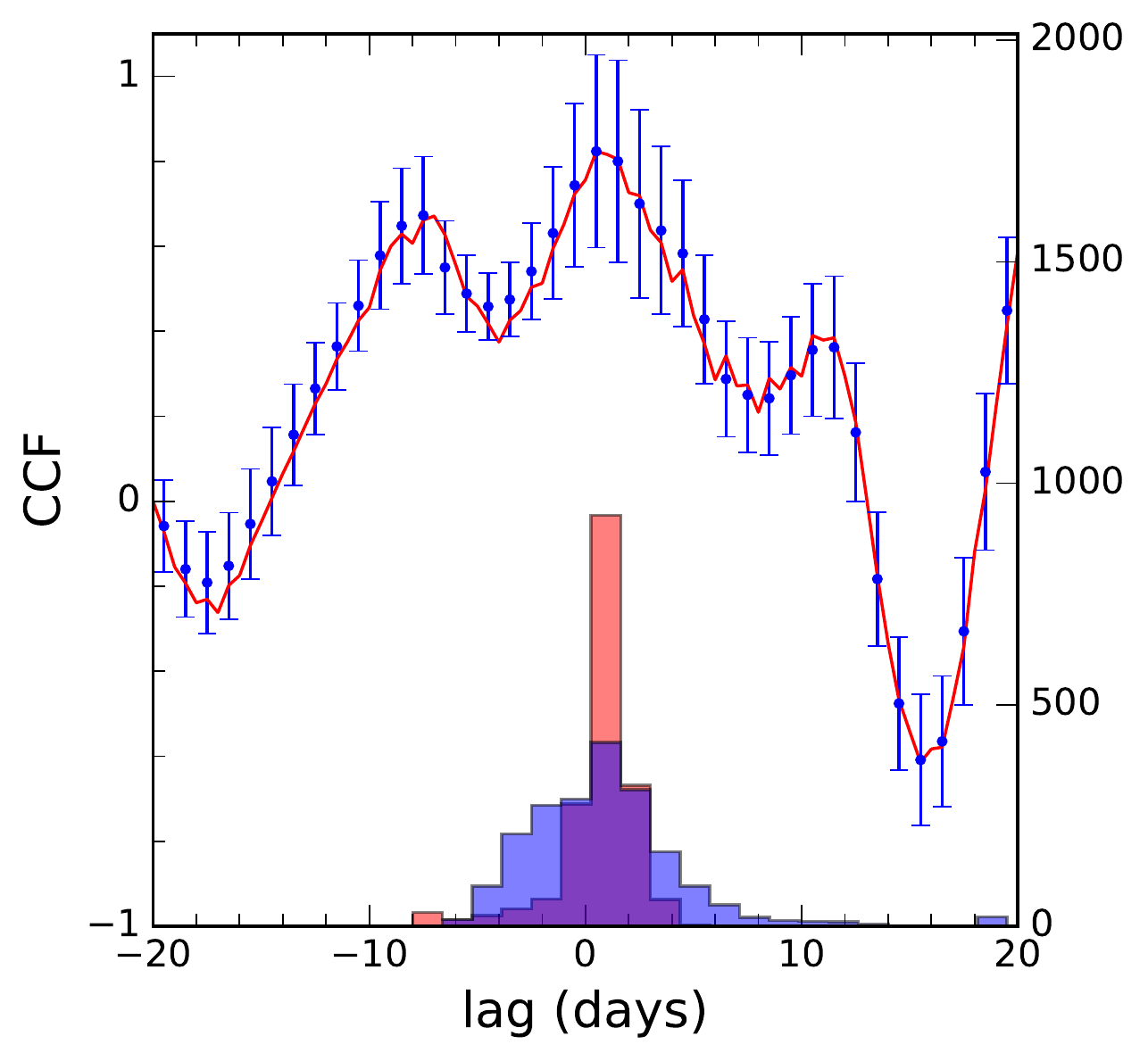}
\caption{Cross-correlation analysis between $\gamma$-ray and  optical 
flux variations during the major flare in epoch B. The solid line is for ICCF and the filled
circles refer to DCF. The histograms in blue and orange show the distribution 
of cross-correlation centroids for ICCF and DCF respectively.}
\label{Fig5}
\end{figure}

\section{Analysis}\label{sec:analysis}

\begin{table}
\caption{Details of the epochs considered for light curves, SED and spectral
 analysis. Here, ID refers to the epochs. The $\gamma$-ray fluxes
are in units of 10$^{-6}$ ph cm$^{-2}$ s$^{-1}$ and the optical fluxes are 
in units of 10$^{-11}$ erg cm$^{-2}$ s$^{-1}$}
\begin{tabular}{lccccccccll}
     \hline
       &\multicolumn{2}{c}{MJD} & \multicolumn{2}{c}{Calender date}& \multicolumn{2}{c}{Mean Flux}\\ 
     \cmidrule(lr){2-3} \cmidrule(lr){4-5} \cmidrule(lr){6-7}
ID     &Start&End&Start&End & $\gamma$ &Opt. \\
     \hline
     A & 55122 & 55222 & 18-10-2009 & 26-01-2010 & 4.97 & 1.76 \\
     B & 55460 & 55560 & 21-09-2010 & 30-12-2010 & 11.7 & 2.81 \\
     C & 55650 & 55750 & 30-03-2011 & 08-07-2011 & 0.54 & 0.68 \\
     D & 56510 & 56610 & 06-08-2013 & 14-11-2013 & 0.99 & 1.67 \\
     E & 56780 & 56880 & 03-05-2014 & 11-08-2014 & 3.75 & 3.23 \\
     
     \hline     
     \end{tabular}
     \label{Table:epochs}
\end{table}

\subsection{Multi-wavelength light curves}
The multi-wavelength light curves that include $\gamma$-ray, X-ray, UV, 
optical and IR along with the polarization measurements from 2008 August 
to 2017 February (MJD: 54500-57800) are shown in Fig \ref{Fig1}. From 
Fig \ref{Fig1} it is evident that 3C 454.3 has gone through both quiescent and active phases during the period MJD 54500 $-$ 57800. During this 
period, we identified four time intervals during which large optical flares 
were seen. They are denoted by epochs A,B,D and E and cover the 
period MJD 55122-55222 (Epoch A), MJD 55460-55560 (Epoch B), 
MJD 56510-56610 (Epoch D) and MJD 56780-56880 (Epoch E). The above four 
intervals were chosen such that (i) there is a 
gradual increase of the optical brightness at least by  0.5 mag from the 
quiescent level (ii) there is a corresponding declining branch from the peak 
back to the quiescent level  and (iii) the rising and decaying phase (both 
inclusive) lasts for more than 50 days. The peak of the flares are shown as a vertical dashed red line in Fig. \ref{Fig1}. On either side of the red lines are two black vertical lines, having a total duration of 100 days. We also identified a time interval 
denoted as Epoch C and covering the period MJD 55650-55750, where the source 
was at its quiescent state in IR-optical-UV-Xrays and $\gamma$-rays. This quiescent period for a duration of 100 days is indicated by two vertical blue lines in Fig. \ref{Fig1}. The
details of the five epochs that were identified for further analysis 
along with their mean optical and $\gamma$-ray flux levels are given in 
Table \ref{Table:epochs}. Detailed analysis of each of these five epochs are given in the 
following sub-sections. 

\begin{figure*}
\vspace*{-2.0cm}
\hspace*{-2.0cm}\includegraphics[width=1.2\textwidth]{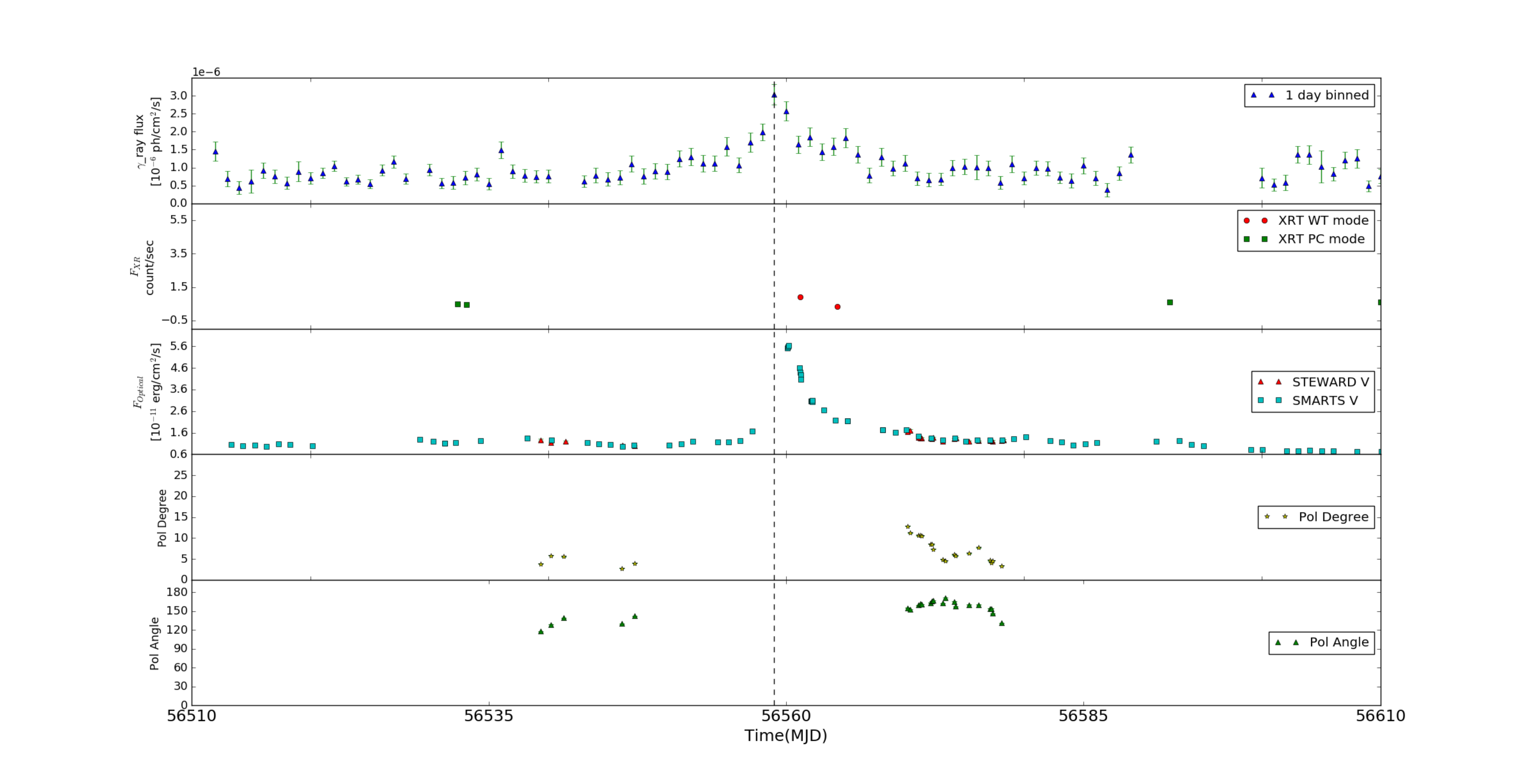}
\caption{Multi-wavelength light curves in $\gamma$-rays, X-ray and optical 
for a period of 100 days during epoch D. The degree of optical polarization
and polarization position angles are shown in the bottom two panels.}
\label{Fig6}
\end{figure*}

\subsubsection{Epoch A (MJD 55122 - 55222)}
An inspection of Fig \ref{Fig1}. indicates  that there is a close correlation 
between IR, optical, UV, X-rays and $\gamma$-rays.  
Optical polarization data though sparse during this period was  not
available during the peak of the flare  making it impossible to comment on the 
nature of the optical polarization during the peak of the $\gamma$-ray flare. 
The multiband light curves covering for a duration of 100 days centered of the peak of epoch A,  
along with the polarization measurements when available are given in 
Fig. \ref{Fig2}.

To check for the presence of any correlation between optical and $\gamma$-ray 
flux variations we cross-correlated the optical and $\gamma$-ray light curves 
using the discrete correlation (DCF) technique of \cite{1988ApJ...333..646E} and the interpolated cross-correlation function (ICCF) technique of \cite{1986ApJ...305..175G,1987ApJS...65....1G} . The 
errors in both DCF and ICCF were obtained by carrying out a Monte Carlo 
analysis that involves both flux randomization and random subset selection  
following the procedures outlined in \cite{2004ApJ...613..682P} . 
The results of the  cross correlation functions analysis are shown  
Fig.\ref{Fig3} both for ICCF and DCF. The lag was determined by the centroid of 
the cross-correlation function which includes all points within 80\% of the
peak of the cross-correlation function.  We found a lag of   
2.2$^{+0.9}_{-0.9}$ days with the $\gamma$-ray leading the optical flux variations. 
This is similar to the lag of about 4 days found between the $\gamma$-ray 
and optical band by \cite{2012AJ....143...23G} on analysis of the 
data for  the time period 2009 November - 2009 December. However,
\cite{2017MNRAS.472..788G} found that the  
optical and $\gamma$-rays are correlated with zero lag  
during the period MJD 55150 - 55200 which is 
within the range analysed here.  During the same period,  
\cite{2017MNRAS.472..788G} found that 
during the declining phase of the $\gamma$-ray flare, the degree of optical 
polarization increased, showing a clear signature of anti correlation between 
$\gamma$-ray flux variation and optical polarization.

\subsubsection{Epoch B (MJD 55460 - 55560)}
During this period, the peak of the optical flare is about two times larger 
than the peak of the optical brightness at epoch A. The $\gamma$-ray 
brightness too peaked at nearly the same time of the optical flare. During 
this epoch, visual inspection indicates close correlation between 
$\gamma$-ray, X-ray, UV optical and IR flux variations. During this period 
a short duration intense flare in the optical was observed superimposed on the 
large optical flare at around MJD 55510. This particular short duration 
optical flare has no corresponding $\gamma$-ray flare (Fig. \ref{Fig4}) 
and is thus a case of an optical flare with no corresponding $\gamma$-ray 
counterpart. At 
the epoch of this short duration optical flare, there is also enhanced optical 
polarization, pointing to a strong correlation between optical flux and
polarization variations. At this time, enhanced flux levels were also seen in 
UV and X-ray bands. This remarkable short duration intense optical flare with no corresponding flare in the gamma-band was also noticed by \cite{2011ApJ...736L..38V}. According to \cite{2011ApJ...736L..38V} this optical flare showed a sharp rise and decay in 48 hours. At the same time, 20\% rise was seen in the X-ray with no change at other wavelengths. During the duration of the large optical flare with the 
peak at MJD 55519, data on the degree of optical polarization is missing to 
make any statement on the correlation or anti-correlation between optical flux 
and polarization variations.   
DCF and ICCF analysis between optical and $\gamma$-ray flux variations, shown 
in Fig. \ref{Fig5} indicate that the time delay between optical and 
$\gamma$-ray flux variations is 0.8$^{+1.1}_{-1.0}$. Thus, during this
major optical flare in epoch B, the optical and $\gamma$-ray flux variations
are correlated with zero lag.

\begin{figure}
\hspace*{-0.2cm}\includegraphics[scale=0.7]{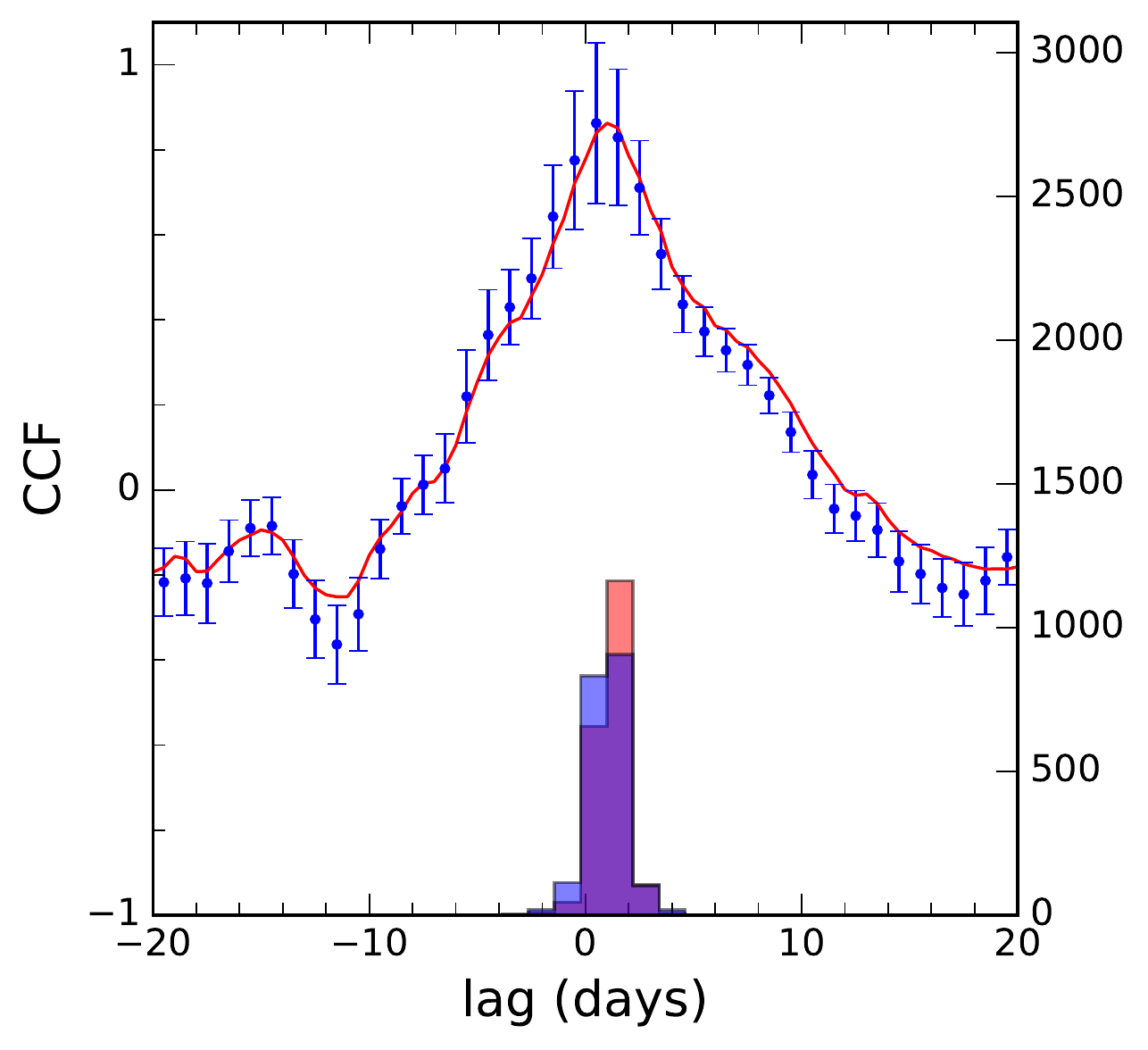}
\caption{Correlation between optical and $\gamma$-ray flux variations
during epoch D. The solid line and filled circles refer to the 
ICCF and DCF respectively. The distribution of the cross-correlation
centroids obtained using ICCF and DCF are shown in blue and orange 
respectively.}
\label{Fig7}
\end{figure}

\begin{figure*}
\vspace*{0.2cm}
\hspace*{-2.0cm}\includegraphics[width=1.2\textwidth]{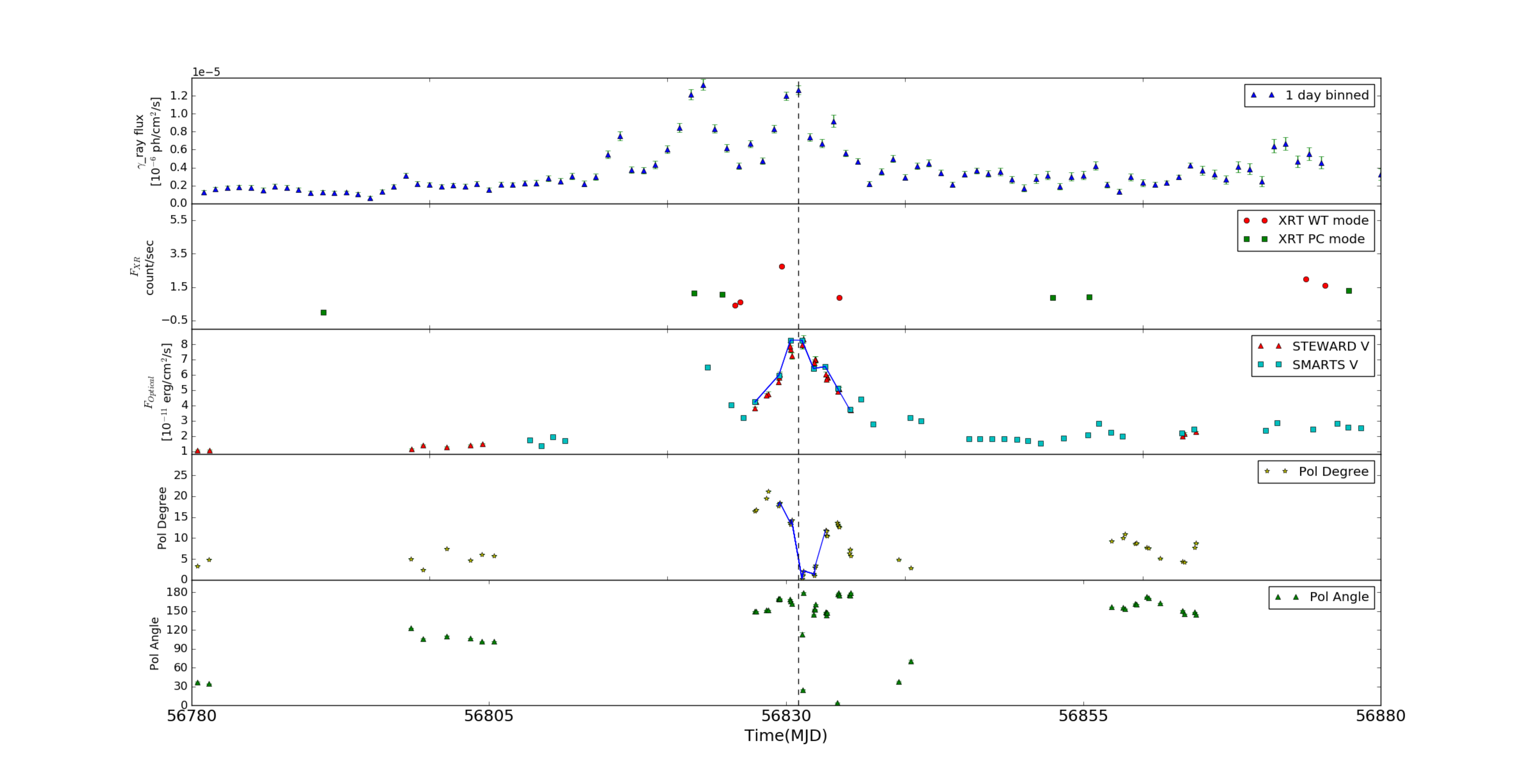}
\caption{\textbf{Multi-wavelength flux and polarization variations during epoch E.}}
\label{Fig8}
\end{figure*}

\subsubsection{Epoch D (MJD 56510 - 56610)}
The optical flux during this epoch has nearly the same amplitude as the 
optical flare at epoch B. Considering the correlation between optical and 
$\gamma$-ray flux variations during both epochs A and B, it is natural to 
expect the $\gamma$-ray flare at epoch D to have similar brightness to that of 
epoch B. However, the source was barely detected in the $\gamma$-ray band 
during this period. This is an indication of an optical flare  with no/weak 
corresponding $\gamma$-ray flare (Fig. \ref{Fig6}). Correlation analysis 
between the optical and $\gamma$-ray light curves during this epoch 
gives a time delay of 1.0$^{+0.7}_{-0.5}$ days. This shows that the 
optical and the very weak $\gamma$-ray variations are correlated with
1 day lag.  The results of the cross-correlation function analysis are
shown in Fig. \ref{Fig7}. Polarizaion data was not available during
the period of the flare and therefore the correlation if any between optical
flux and polarization variations could not be ascertained.

\begin{figure}
\hspace*{-0.2cm}\includegraphics[scale=0.7]{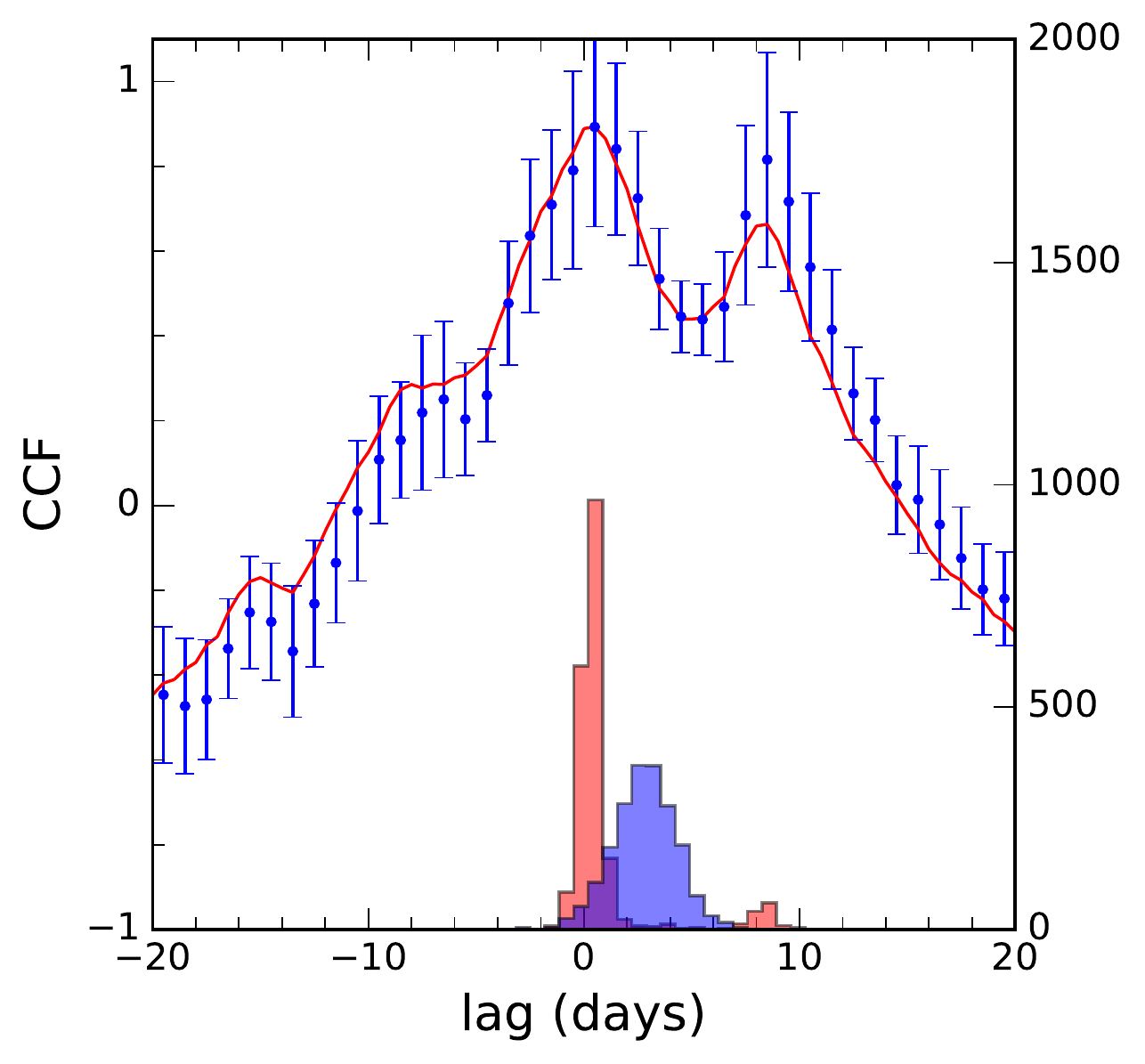}
\caption{Correlation between optical and $\gamma$-ray flux variations
during epoch E. The solid line and filled circles refer to the 
ICCF and DCF respectively. The distribution of the cross-correlation
centroids obtained using ICCF and DCF are shown in blue and orange 
respectively.}
\label{Fig9}
\end{figure}

\subsubsection{ Epoch E (MJD 56780 - 56880)}
During this epoch the optical flare has a peak brightness similar to that of 
the optical flare at epoch B, but the source has minor $\gamma$-ray flare during this epoch.  
This same period was also independently analyzed by \citep{2017MNRAS.464.2046K} for correlation
between $\gamma$-ray and optical flux variations. They find no lag between 
optical and $\gamma$-ray flux variations during the period overlapping the 
duration of epoch E. Our Correlation analysis 
between the optical and $\gamma$-ray light curves during this epoch 
gives a time delay of 0.3$^{+0.7}_{-0.5}$ days.Correlation analysis for this epoch shown in figure \ref{Fig9}. We noticed an interesting feature by careful examination 
of the optical total flux and polarization variations shown in Fig \ref{Fig8}. 
The degree of optical polarization is anticorrelated with the optical flare 
both during the rising phase and the decaying phase of the flare. Though such 
anti-correlations between optical flux and polarization variations were 
known before in the blazar BL Lac 
\citep{2014ApJ...781L...4G} and 3C 454.3 \citep{2017MNRAS.472..788G}, 
we noticed anticorrelation between optical flux and polarization variations 
both during the rising part of the flare as well as the decaying part of the 
flare.

\begin{figure*}
\begin{center}$
\begin{array}{lll}
\includegraphics[width=90mm]{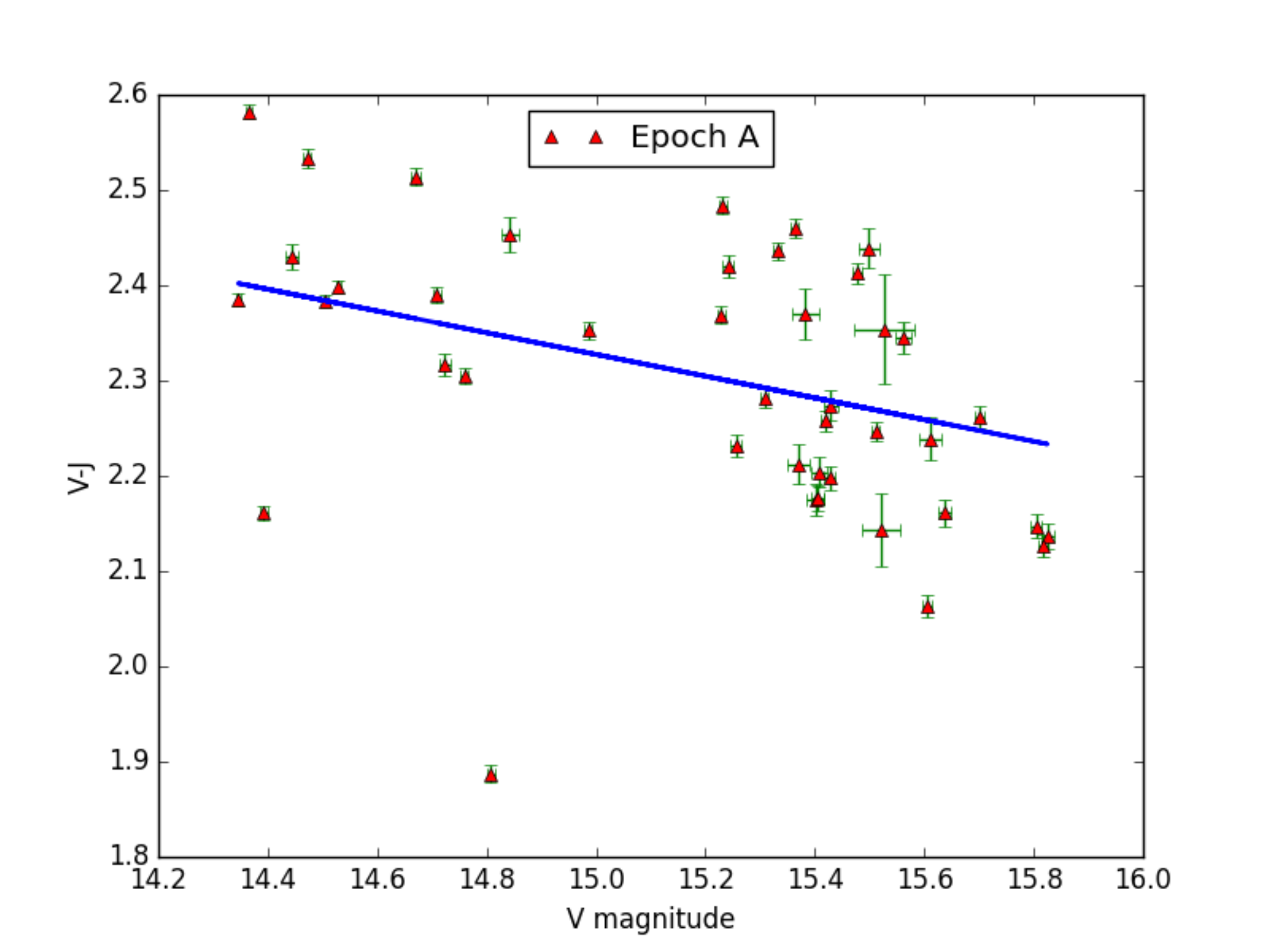}&
\includegraphics[width=90mm]{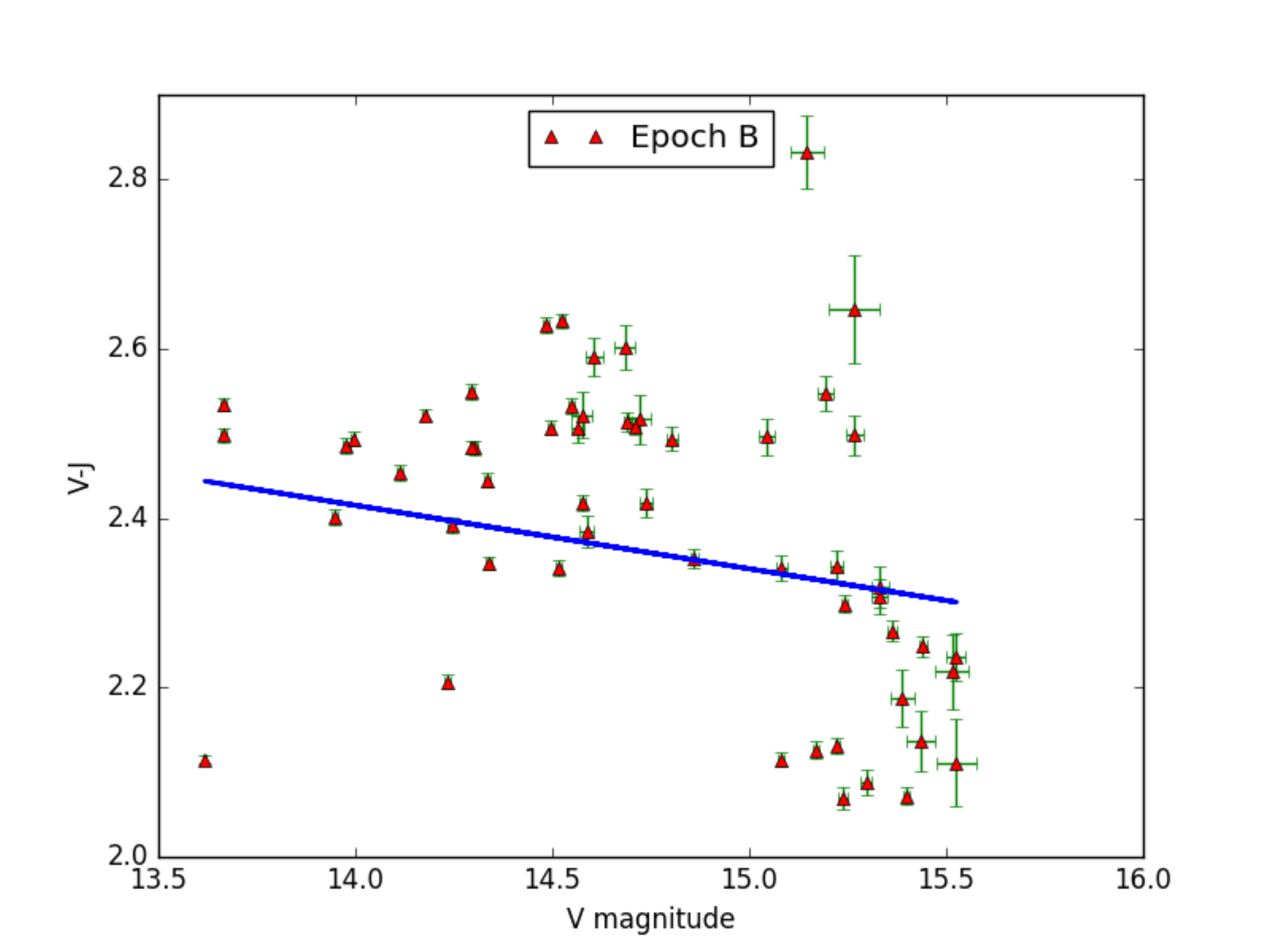}
\end{array}$
\end{center}

\begin{center}$
\begin{array}{rr}
\includegraphics[width=90mm]{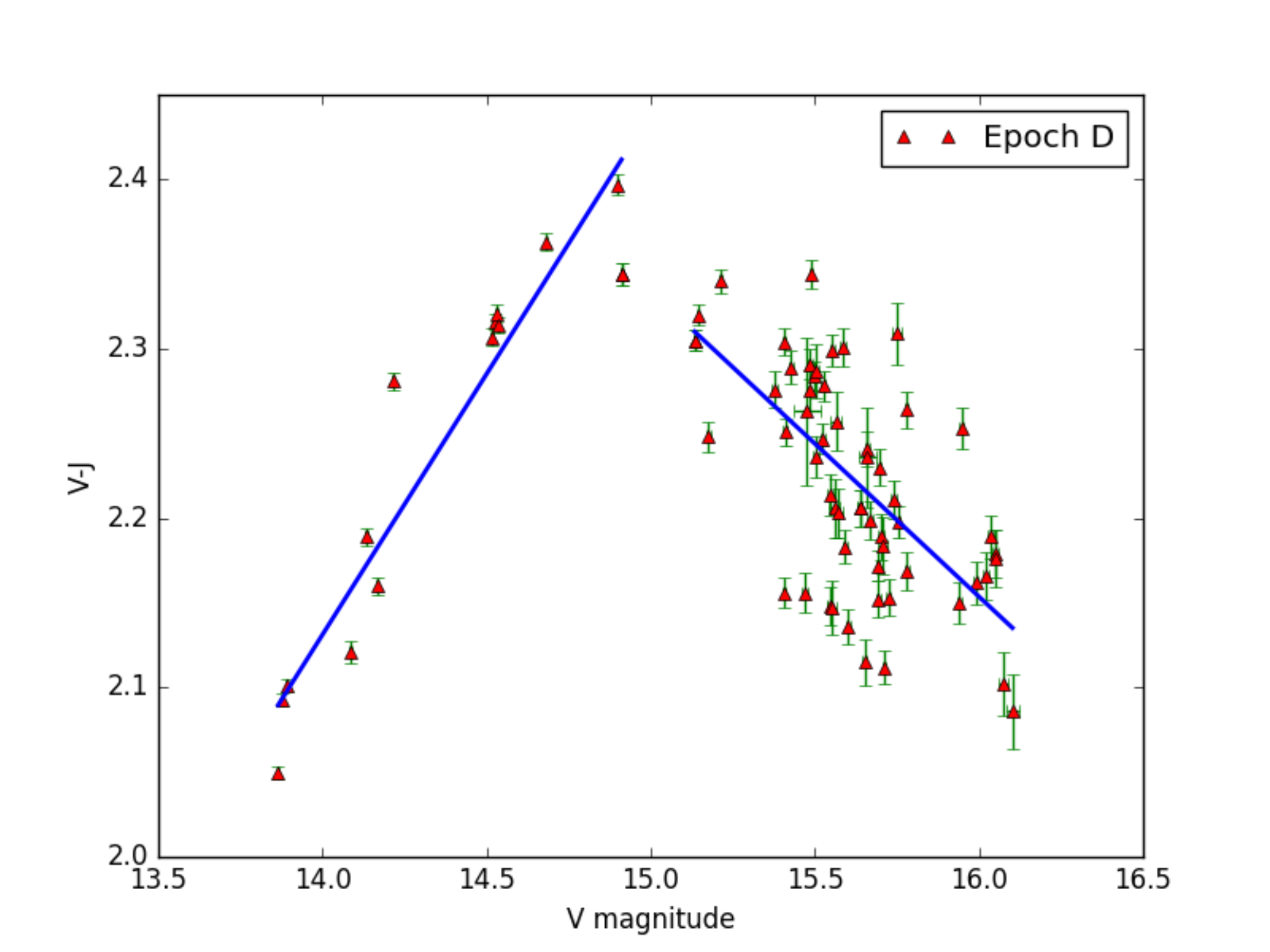}&
\includegraphics[width=90mm]{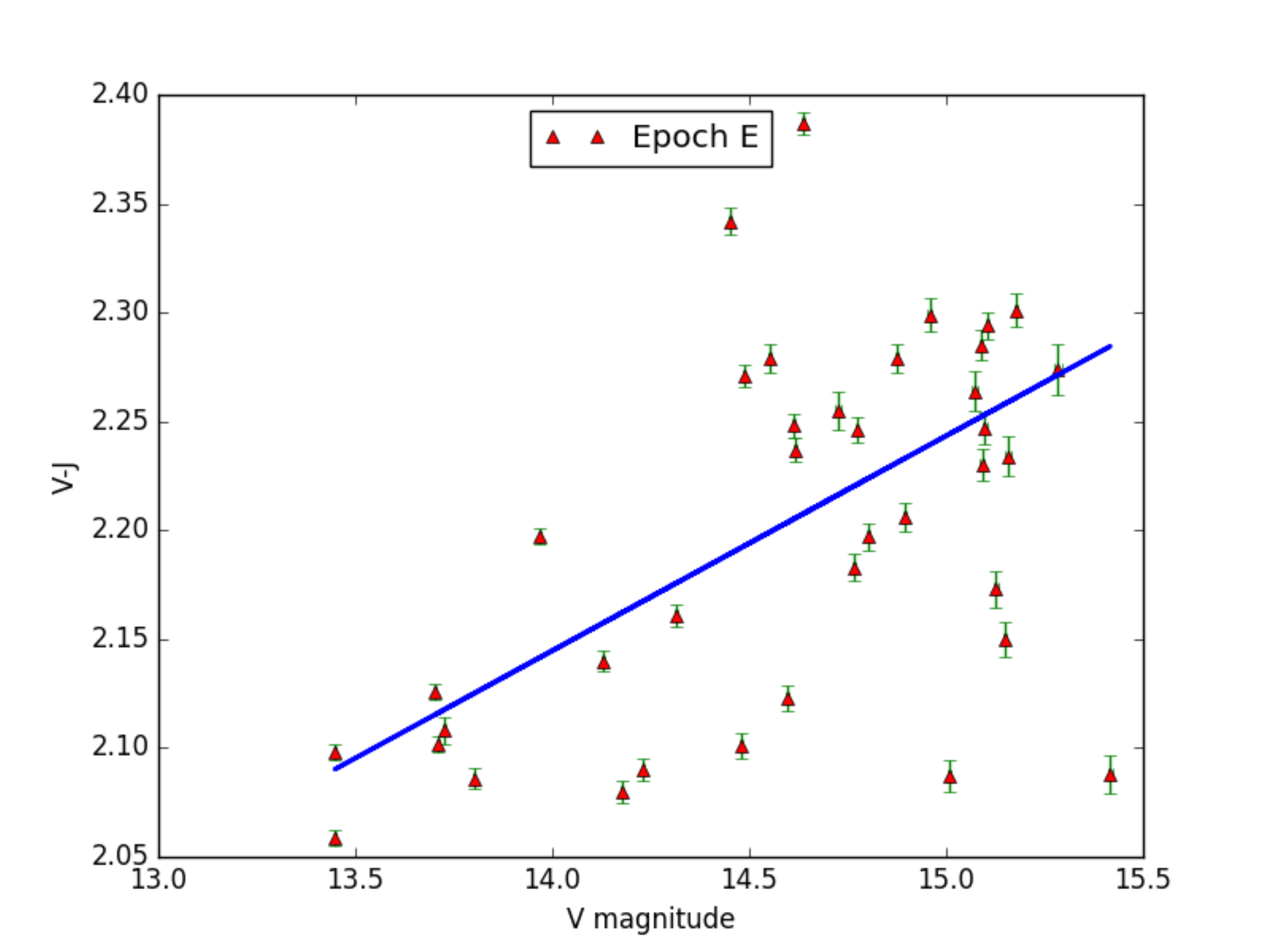}
\end{array}$
\end{center}

\caption{Colour-magnitude relations for  epochs  A, B, D and E}
\label{Fig10}
\end{figure*}

\subsection{Spectral variations}
To check for any spectral variation in the optical/IR bands, we looked for variation in the V-J band
colour against the V-band brightness. This colour variation was analyzed for the epochs A, B, D and E.
During epoch A and B, the source showed a "redder when brighter (RWB)" behavior. During epochs E, a bluer
when brighter behavior was observed. During epoch D, we observed a complex behavior. Upto a V-band
brightness of around 15 mag, the source showed a "bluer when brighter" behavior, but for optical
brightness fainter than 15.0 mag, a "redder when brighter" behavior was observed. The colour magnitude
diagram for all the four epochs are shown in Fig. \ref{Fig10}. The spectral variations shown by the source is 
thus complex. From studies on the optical-IR colour-magnitude diagram, it is known that
FSRQs in general show a RWB trend, which is 
attributed to them having luminous accretion disk \citep{2012ApJ...756...13B,2006A&A...450...39G}. The observed optical emission is a combination of thermal
blue emission from the accretion disk and non-thermal red emission from the 
jet. As the source gets brighter, the non-thermal emission has a more 
dominant contribution to the total flux giving rise to the RWB behavior
\citep{2012ApJ...756...13B}. During epochs A and B, there is a trend of the object to become RWB, irrespective of its optical
brightness. The optical flares dominated by synchrotron emission processes during A and B have corresponding $\gamma$-ray flares that are produced by EC processes. However, during epochs D and E, the colour variations were found to depend on the optical brightness. During the epochs when this complex spectral behavior was noticed, the source showed optical/IR flare with no or weak
corresponding flare in the $\gamma$-ray band. The source showed a much larger amplitude of variability in the optical/IR bands, while in the $\gamma$-ray band it was either faint or below the detection limit of {\it Fermi}. This definitely points to some complex physical changes and could be due to a combination of changes in the bulk Lorentz factor, electron energy density
and magnetic field as seen from our SED modeling of the multiband data.

\subsection{$\gamma$-ray spectra}
The shape of the $\gamma$-ray spectrum can provide evidence on the intrinsic 
distribution of electrons involved in the $\gamma$-ray emission processes that 
might involve acceleration and cooling processes. For all the five intervals 
identified above, we generated the $\gamma$-ray spectra and fitted them with 
two models, namely a simple power law (PL) model and a log parabola (LP) model. The PL model has the form
\begin{equation}
dN(E)/dE=N_{\circ}(E/E_{\circ})^{-\Gamma}
\end{equation}
where $N_{\circ}$ is normalization of the energy spectrum and $E_{\circ}$ = 300MeV, which is constant for all SEDs.

The LP model is defined as below following \cite{2012ApJS..199...31N}  

\begin{equation}
dN(E)/dE=N_{\circ}(E/E_{\circ})^{-\alpha-\beta ln(E/E_{\circ})}
\end{equation}
here, dN/dE is the number of photons cm$^{-2}$ s$^{-1}$ MeV$^{-1}$,  $\alpha$ 
is photon index at $E_{\circ}$, $\beta$ is the curvature index, E is the 
$\gamma$-ray photon energy, N$_{\circ}$ and E$_{\circ}$ are the 
normalization and scaling factor of the energy spectrum respectively.

\begin{figure*}
\begin{center}$
\begin{array}{lll}
\includegraphics[width=90mm]{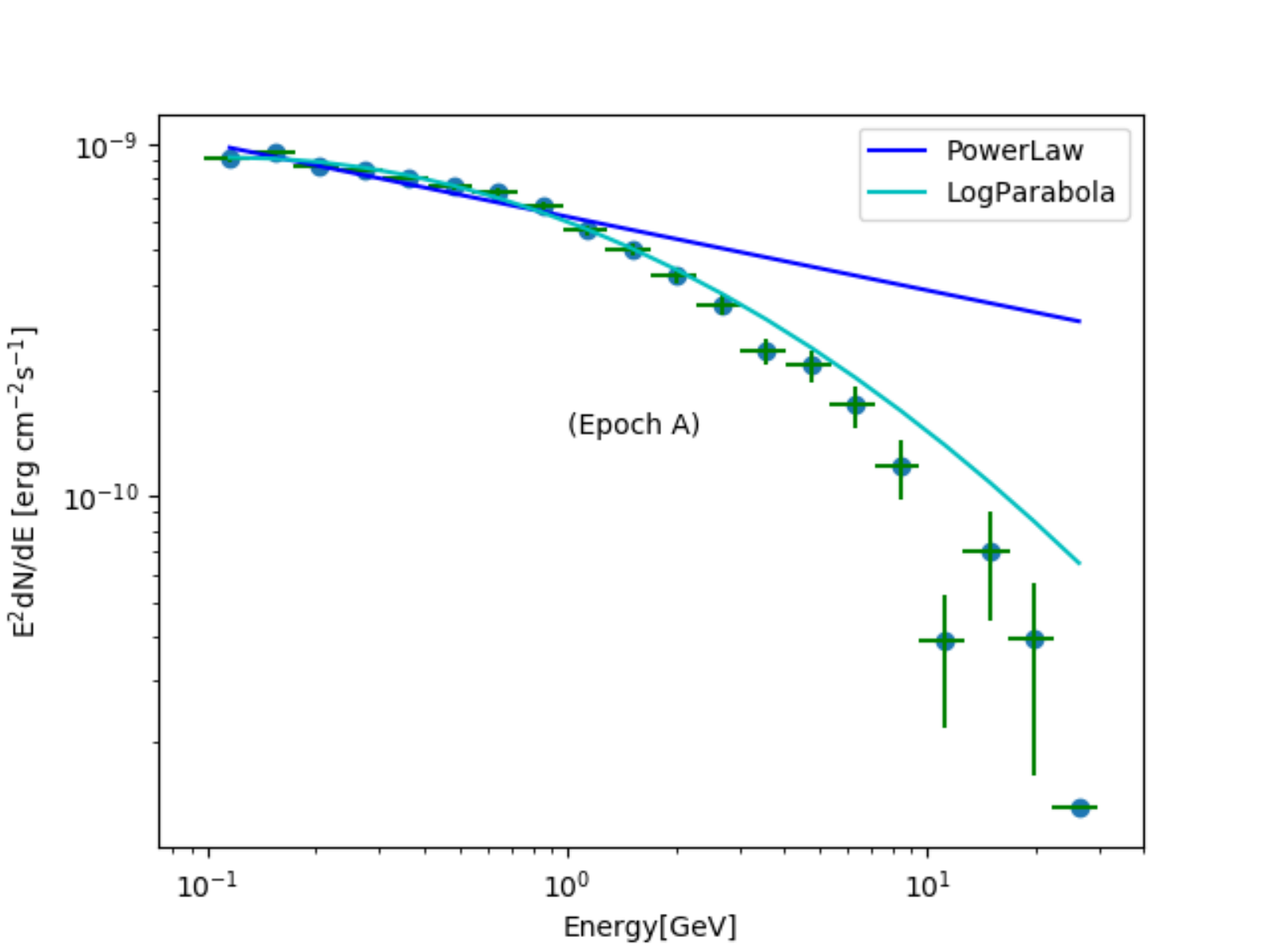}&
\includegraphics[width=90mm]{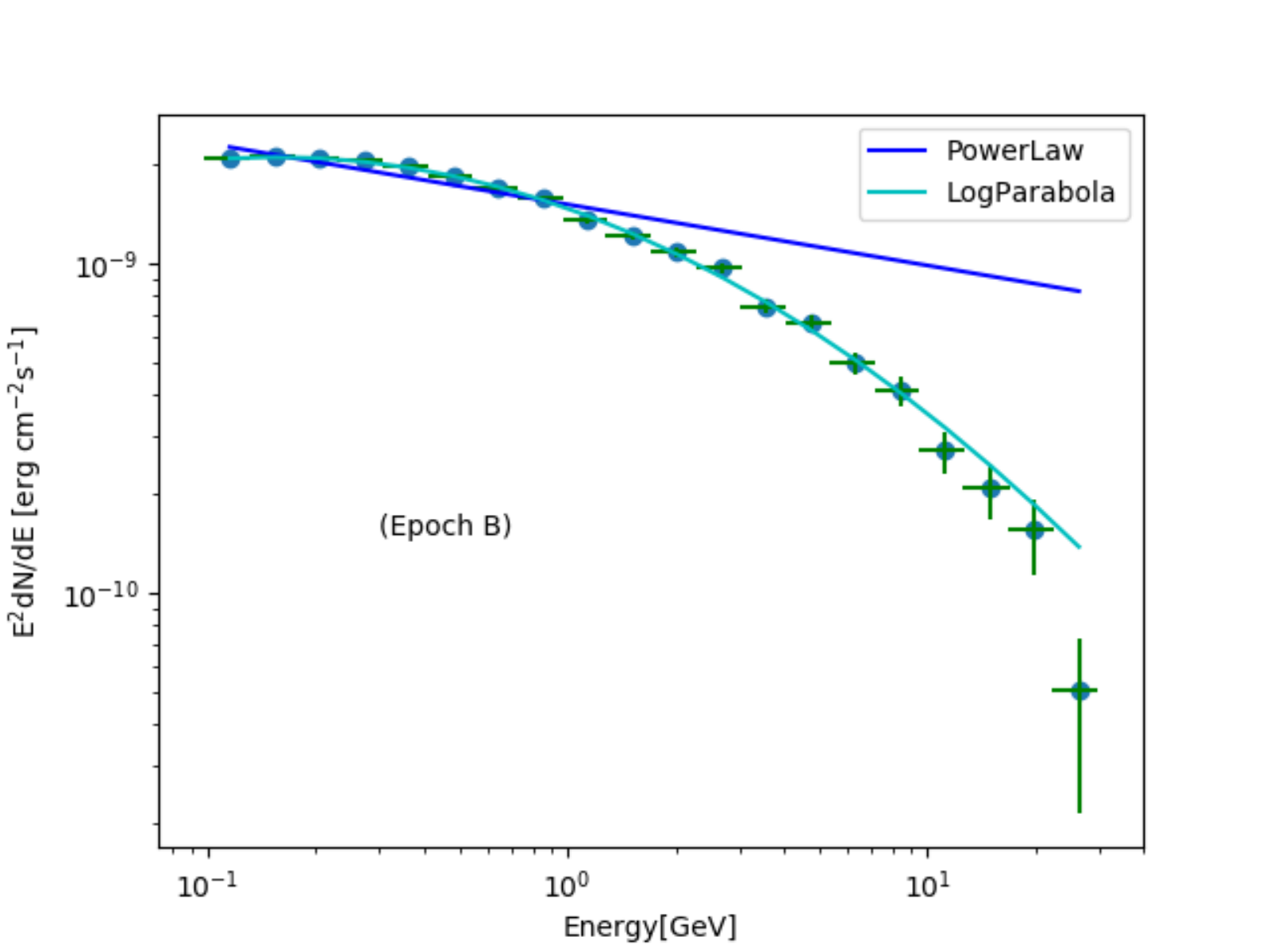}
\end{array}$
\end{center}

\begin{center}$
\begin{array}{rr}
\includegraphics[width=90mm]{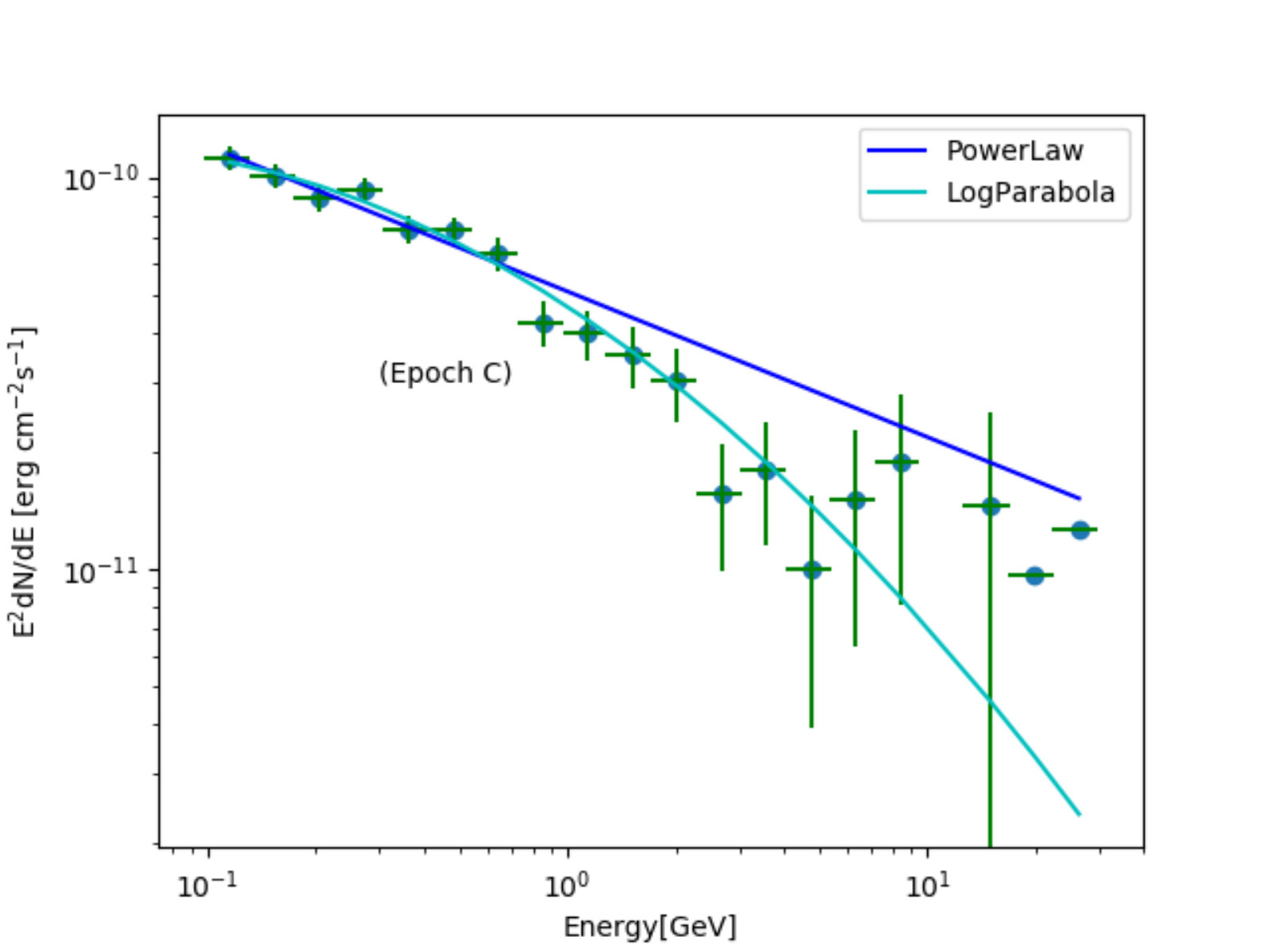}&
\includegraphics[width=90mm]{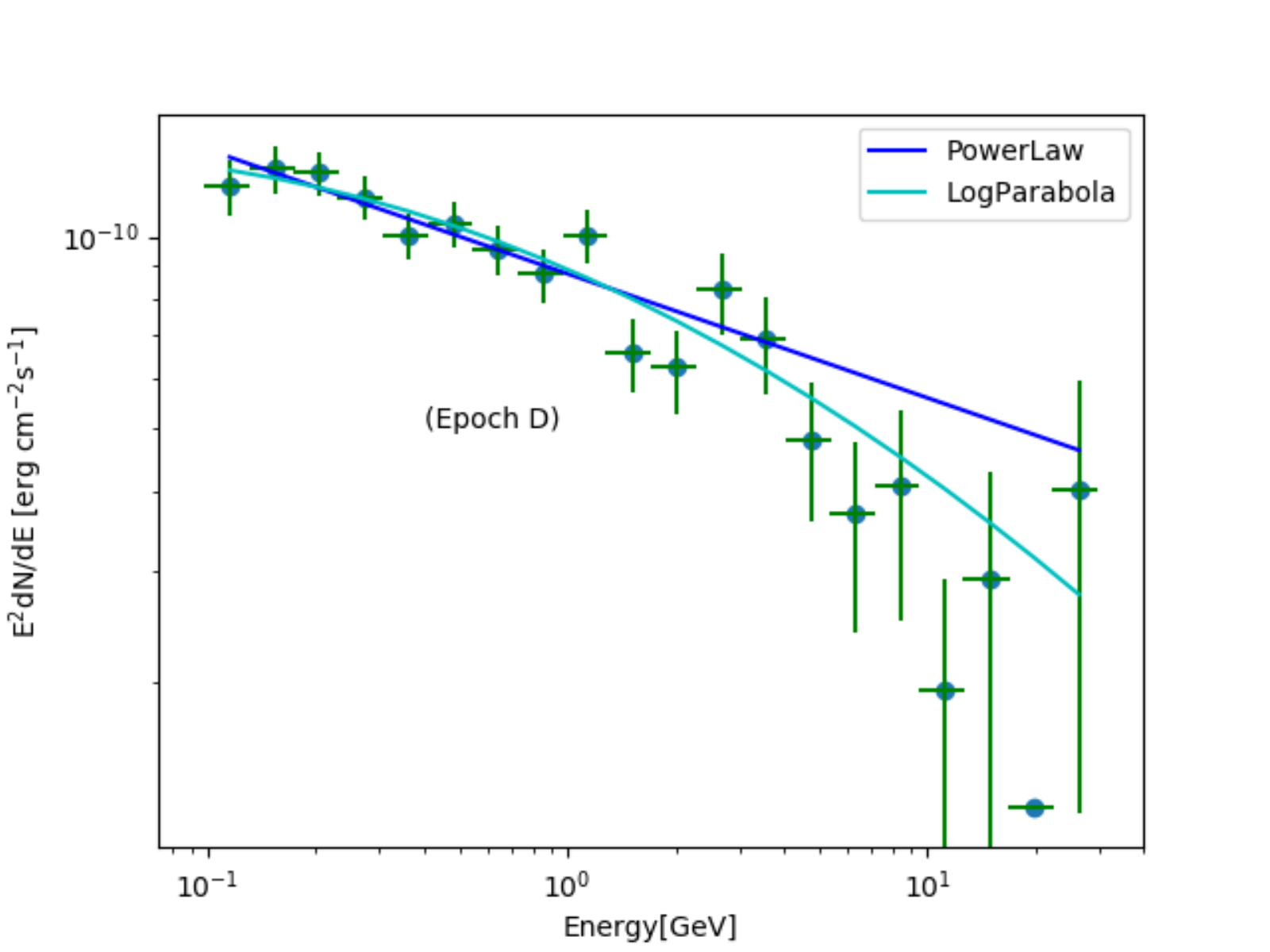}
\end{array}$
\end{center}

\begin{center}$
\begin{array}{rr}
\includegraphics[width=100mm]{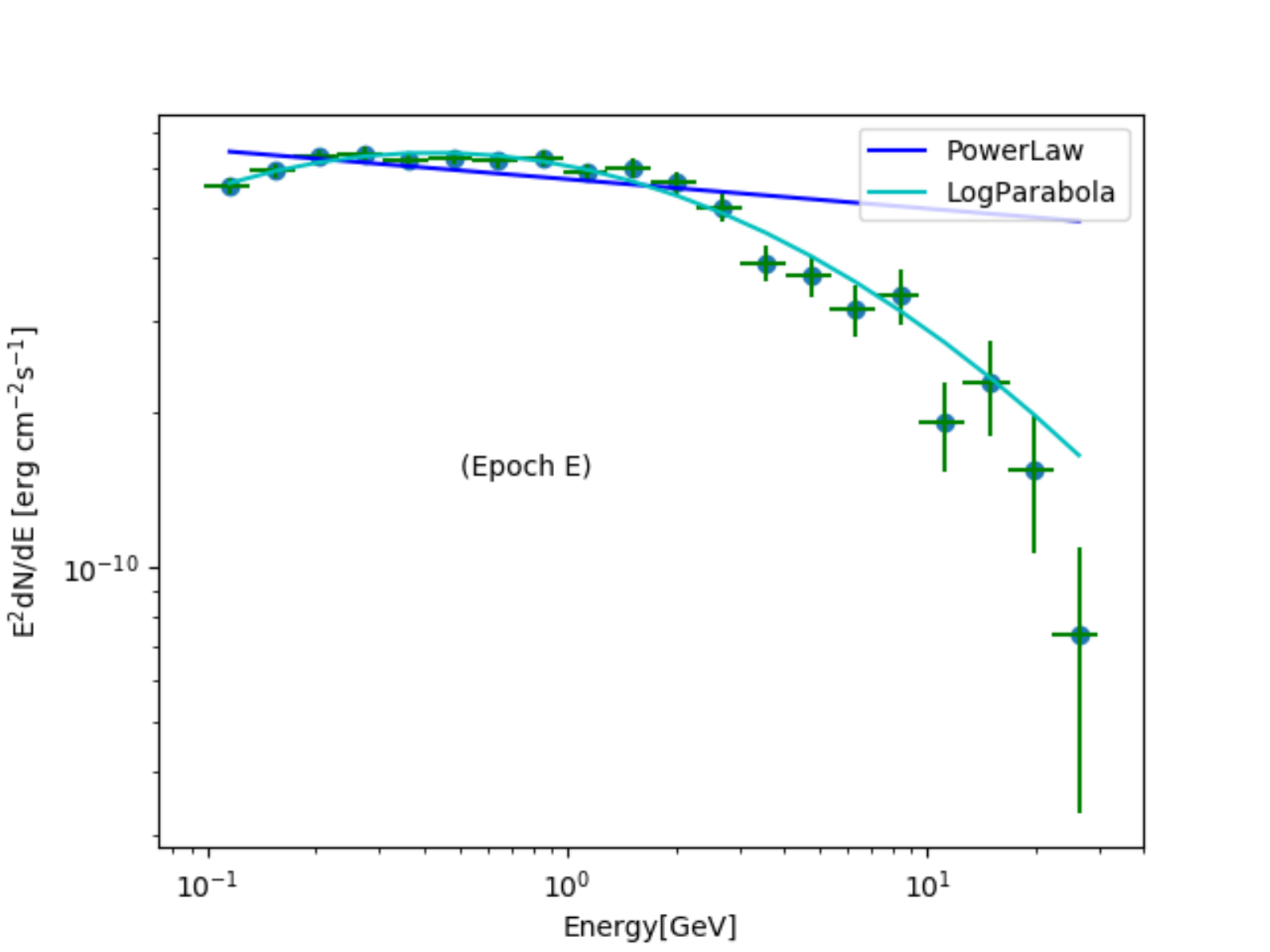}
\end{array}$
\end{center}
\caption{Simple power law and Log parabola fits to the 
$\gamma$-ray spectra of 3C 454.3 during epochs A, B, C, D, and  E}
\label{Fig11}
\end{figure*}

We used Maximum Likelihood estimator gtlike for spectral analysis likelihood ratio 
test \citep{1996ApJ...461..396M} to check the PL model (null hypothesis) 
against the LP model(alternative hypothesis). $TS_{curve}$ = 
2(log $L_{LP}$ - log $L_{PL}$) was also calculated \citep{2012ApJS..199...31N}.
The presence of a significant curvature was tested by setting the condition 
$TS_{curve}$ > 16. Gamma-ray spectrum for these five epochs are shown in Figure \ref{Fig11} and the results of the $\gamma$-ray spectral analysis are shown in Table \ref{Table:PL}. On all the five epochs the $\gamma$-ray spectra is well fit with a LP model.

\subsection{Spectral energy distribution modeling}
To characterize the nature of the source during epochs A, B, C, D and E, we 
constructed the broad band spectral energy distribution. For UV, optical and IR, all data points over the 100 day period in each of the epochs were averaged filter wise to get one data point for each filter. However, for X-ray and $\gamma$-rays, all the available data over the 100 day period in each of the epochs was used to construct their average spectra. All the generated SEDs were modeled using the one zone leptonic model of \cite{2012MNRAS.419.1660S}. In this model, the emission region is assumed to be a spherical blob of size $R$ filled with non-thermal electrons following a broken power law distribution
\begin{align} \label{eq:broken}
	N(\gamma)\,d\gamma = \left\{
\begin{array}{ll}
	K\,\gamma^{-p}\,d\gamma&\textrm{for}\quad \mbox {~$\gamma_{\rm min}<\gamma<\gamma_b$~} \\
	K\,\gamma_b^{q-p}\gamma^{-q}\,d\gamma&\textrm{for}\quad \mbox {~$\gamma_b<\gamma<\gamma_{\rm max}$~}
\end{array}
\right.
\end{align}
where, $\gamma$ is the electron Lorentz factor and, $p$ and $q$ are the low and high energy power-law indices with $\gamma_b$ the Lorentz factor corresponding to the break energy. The emission region is permeated with a tangled magnetic field $B$ and move down the jet with a bulk Lorentz factor $\Gamma$. The broadband SEDs are modelled using synchrotron, SSC and EC emission 
mechanisms. This model was added as a local model in XSPEC \citep{1996ASPC..101...17A} and the source parameters were obtained through $\chi^{2}$ minimization \citep{2018RAA....18...35S}. The observed spectrum is mainly governed by $10$ free parameters with $4$ of them governing the electron distribution namely, $p$, $q$, $\gamma_b$ and the electron energy density $U_e$. The rest of the $6$ parameters are, $B$, $R$, $\Gamma$, jet viewing angle $\theta$, the temperature of the external thermal photon field $T$ and the fraction of the external photons taking part in the EC process, $f$. To account for the model related uncertainties, we added 12\% systematic error evenly over the entire data. For SED model fits, corrections due to galactic absorption were applied to the IR, optical\footnote{http://ned.ipac.caltech.edu}, UV \citep{2011A&A...534A..87R} and X-ray data points.

\begin{figure*}
\begin{center}$
\begin{array}{lll}
\includegraphics[width=95mm,height=75mm]{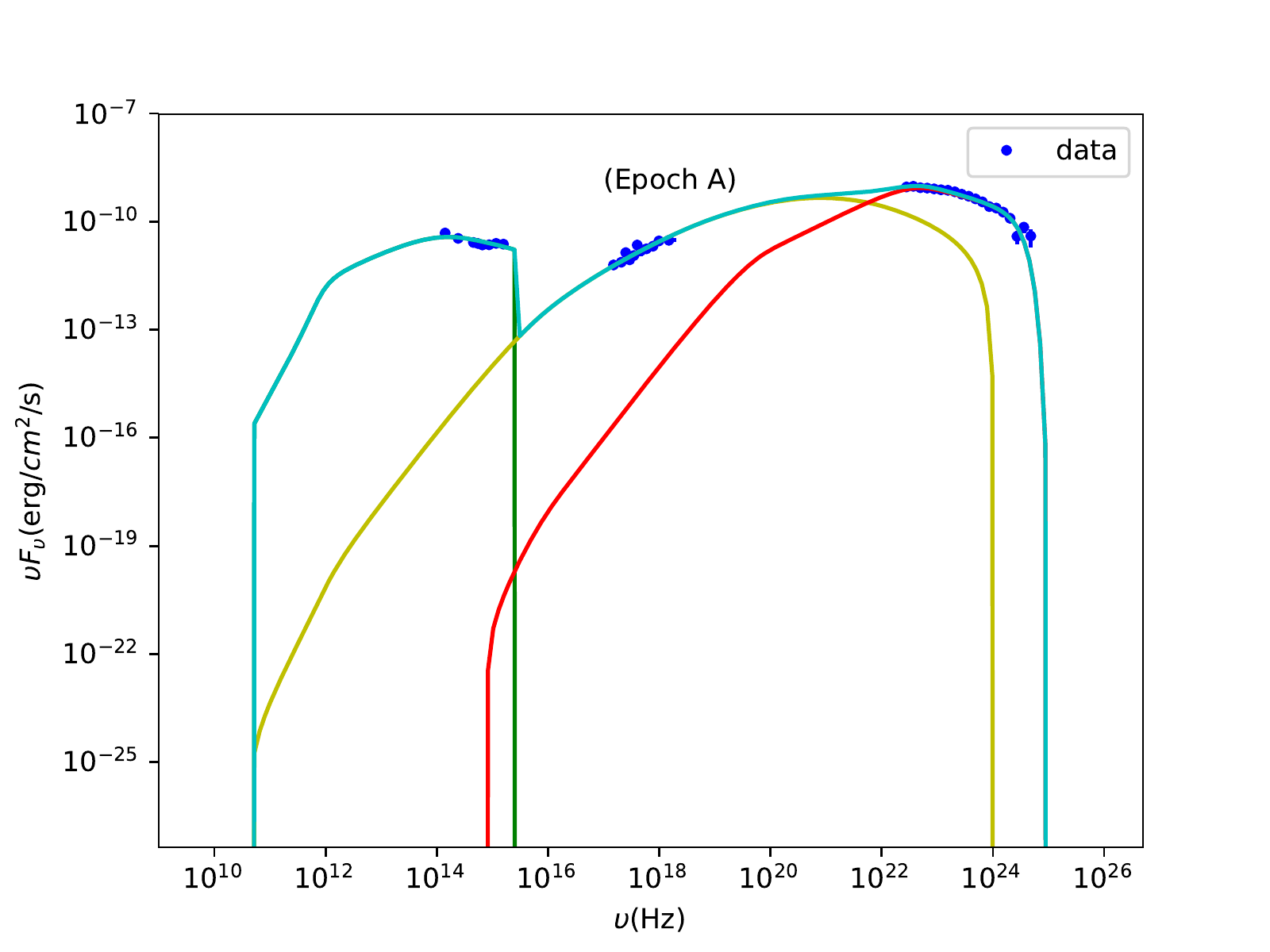}&
\includegraphics[width=95mm,height=75mm]{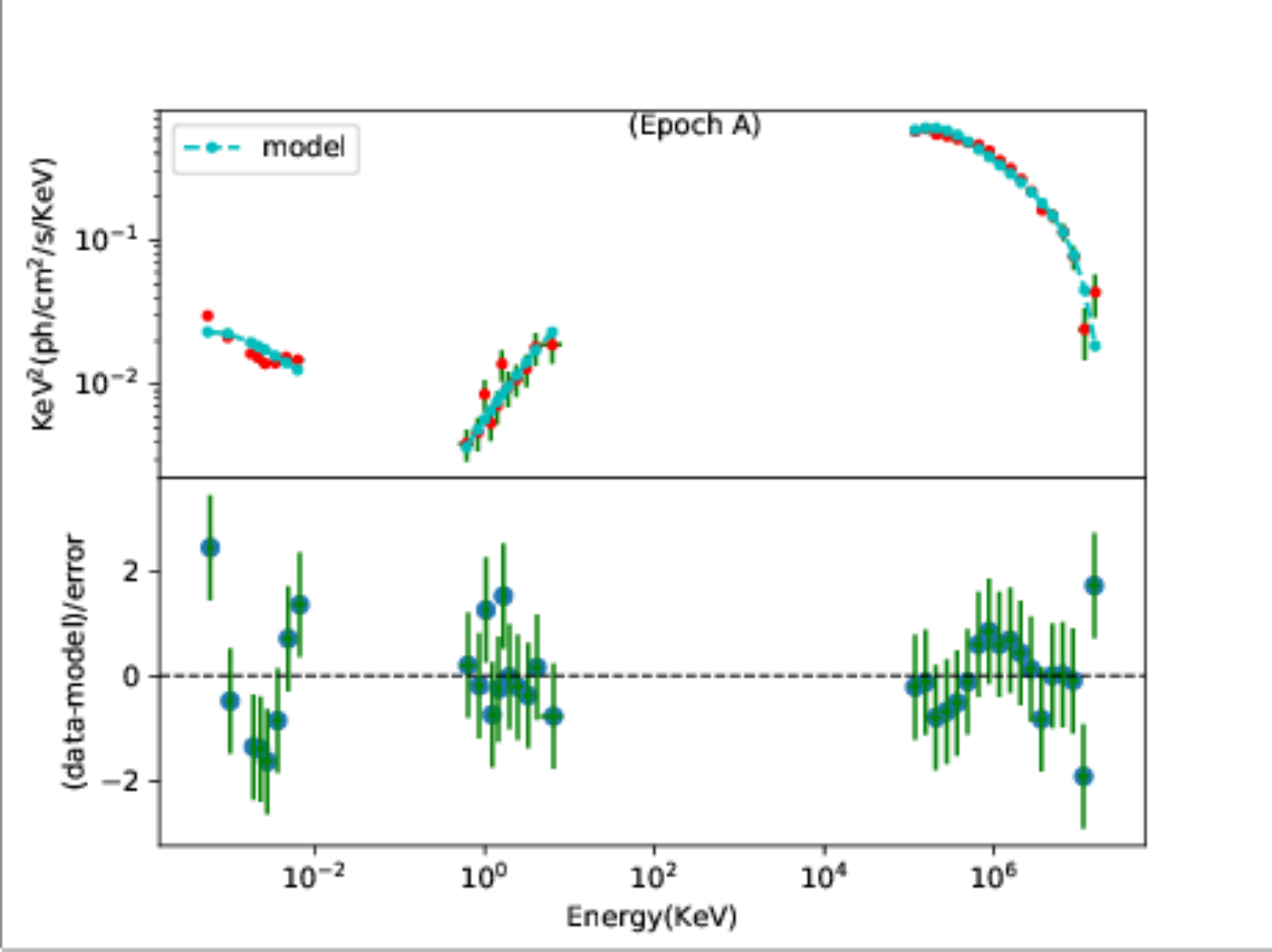}
\end{array}$
\end{center}

\begin{center}$
\begin{array}{rr}
\includegraphics[width=95mm,height=75mm]{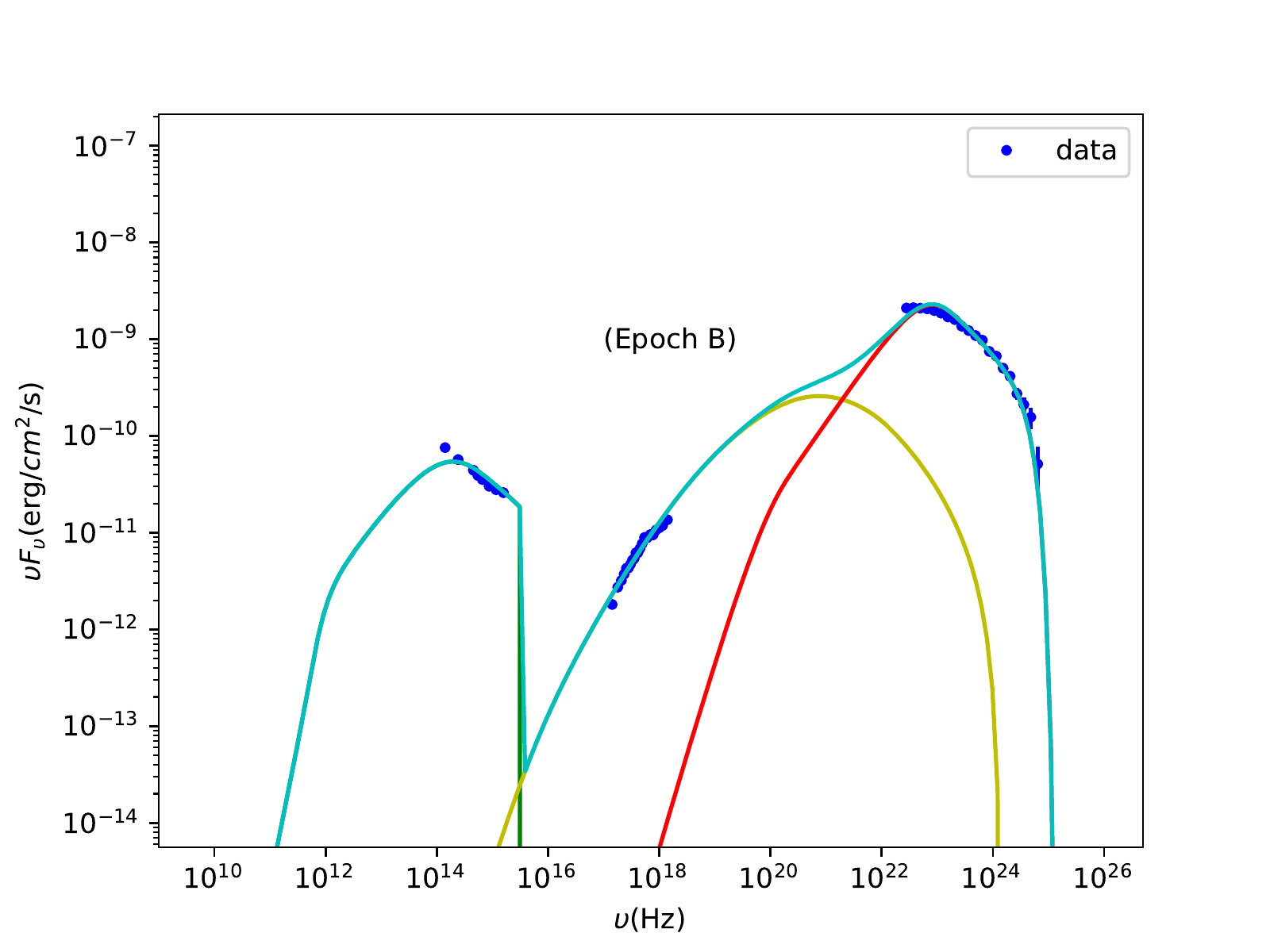}&
\includegraphics[width=95mm,height=75mm]{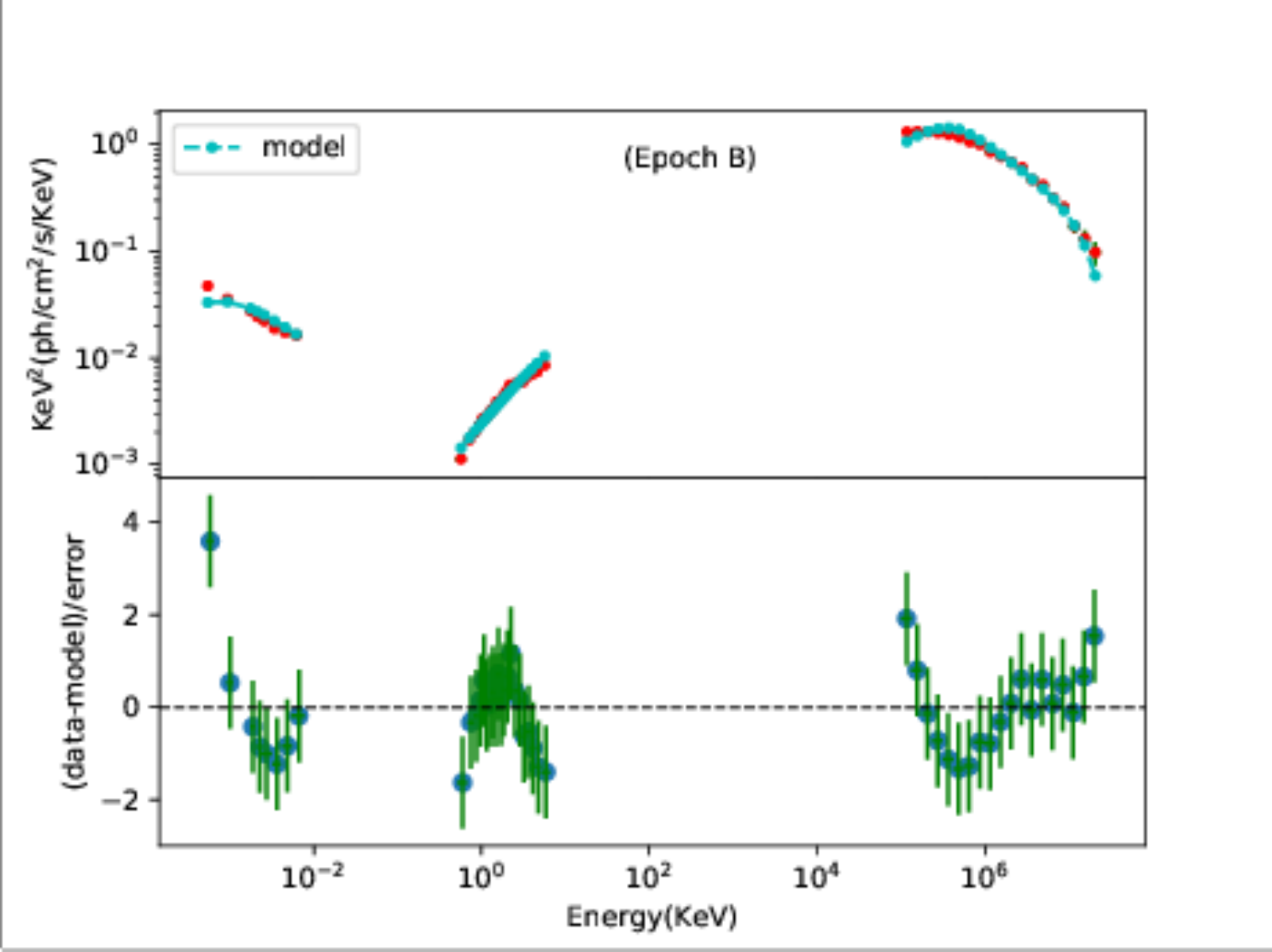}
\end{array}$
\end{center}

\begin{center}$
\begin{array}{rr}
\includegraphics[width=95mm,height=75mm]{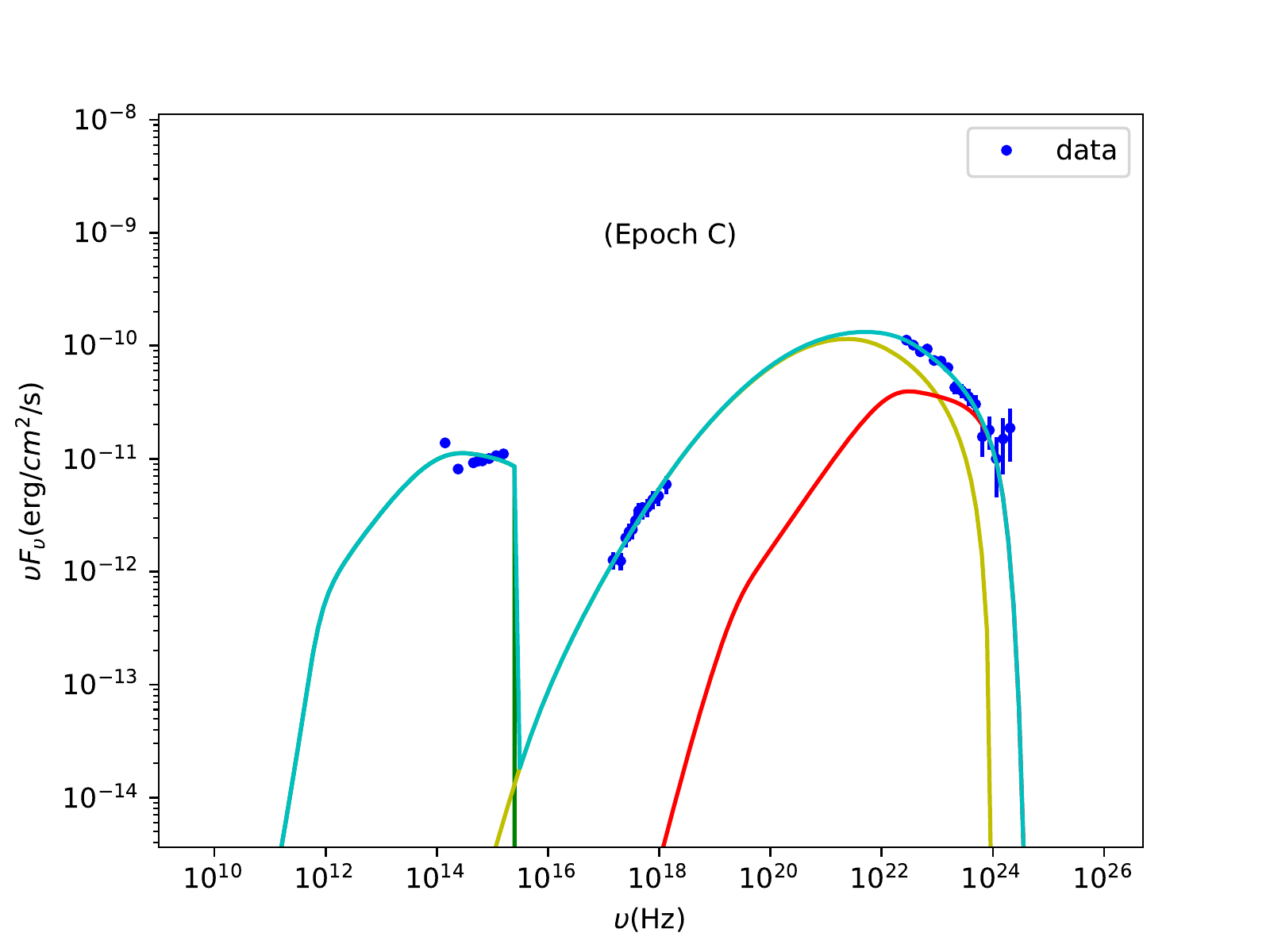}&
\includegraphics[width=95mm,height=75mm]{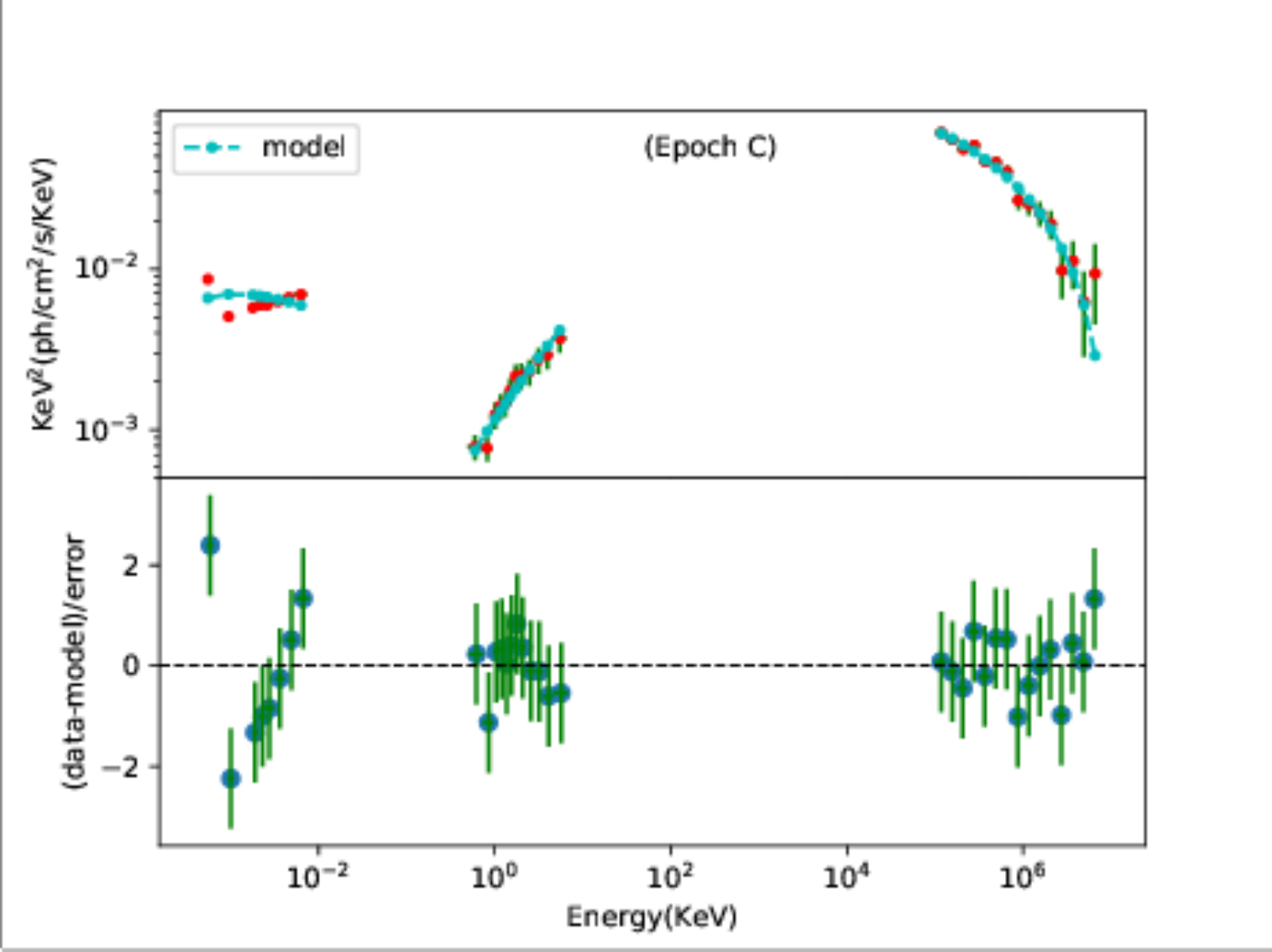}
\end{array}$
\end{center}
\vspace*{-0.5cm}\caption{Broad band spectral energy distribution along with the 
one zone leptonic emission model fits for epochs  A, B and  C. In the left
panels the green line refers to the synchrotron model, the yellow line refers to the SSC process and the red line refers to the
EC process. The cyan line is the sum of all the components. In the right hand
panels for each epoch the first panel shows the fitting of the model to the
data carried out in XSPEC and the second panel shows the residuals.}
\label{Fig12}
\end{figure*}

\begin{figure*}
\begin{center}$
\begin{array}{rr}
\includegraphics[width=95mm,height=75mm]{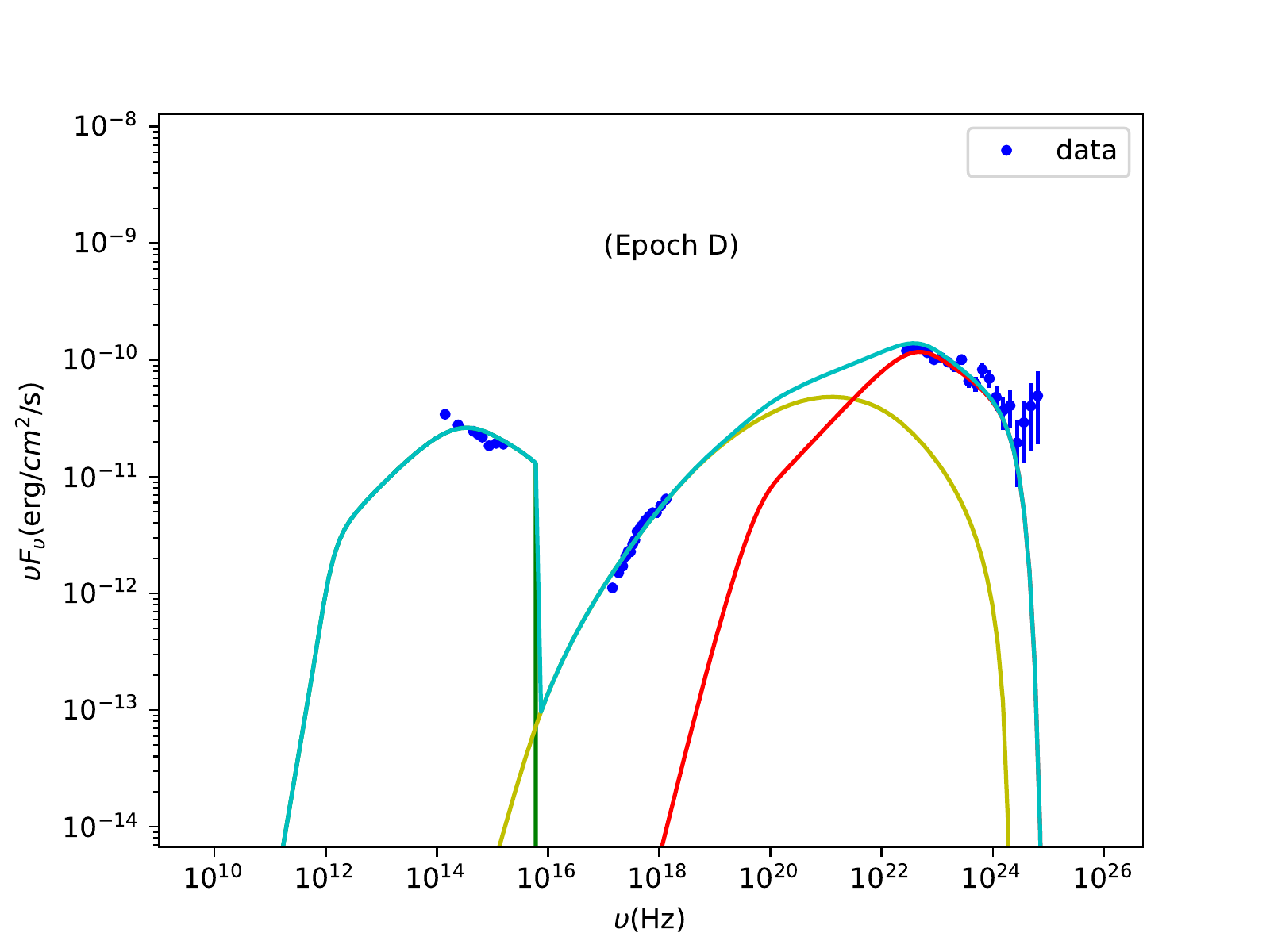}&
\includegraphics[width=95mm,height=75mm]{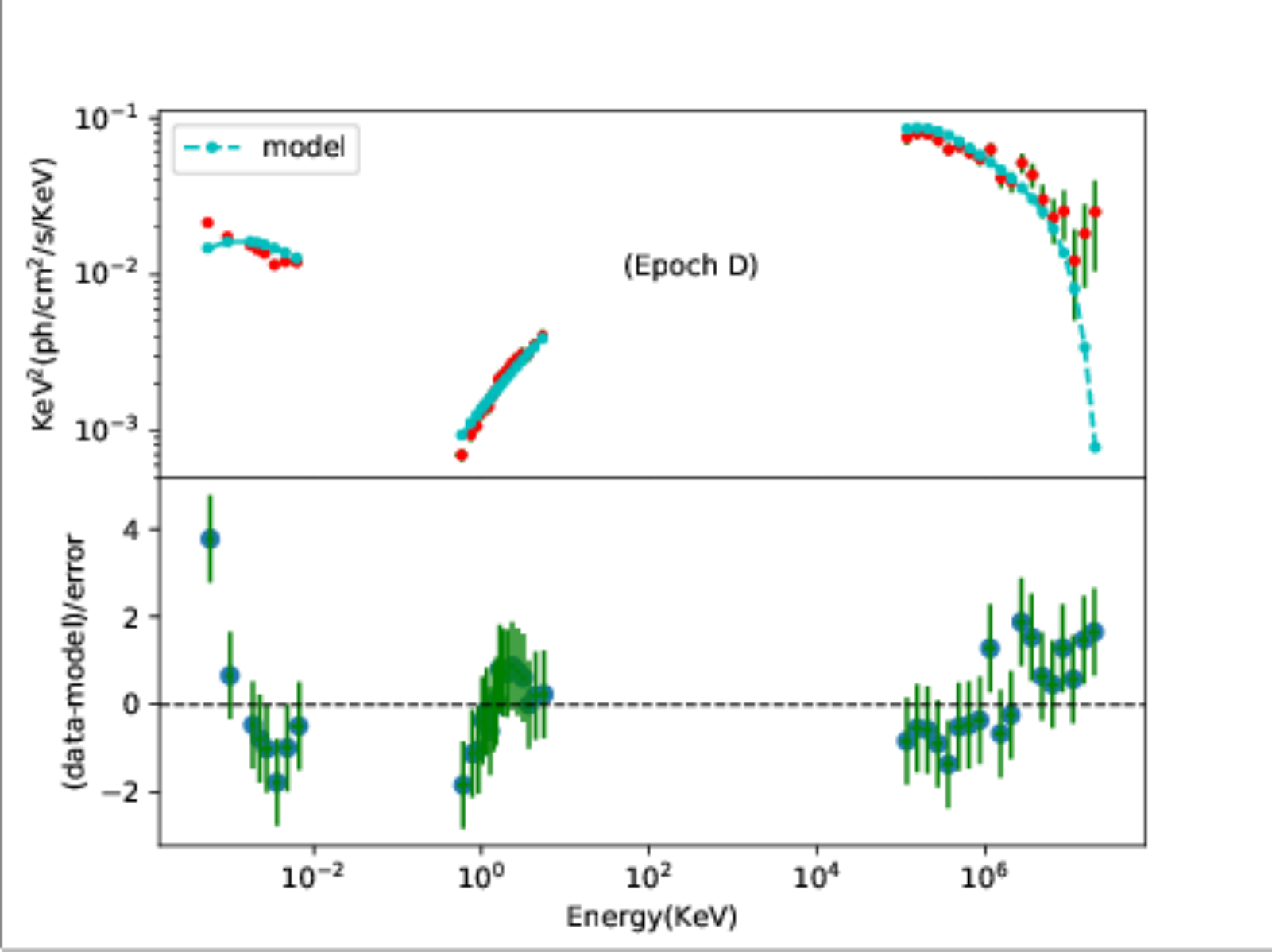}
\end{array}$
\end{center}

\begin{center}$
\begin{array}{rr}
\includegraphics[width=95mm,height=75mm]{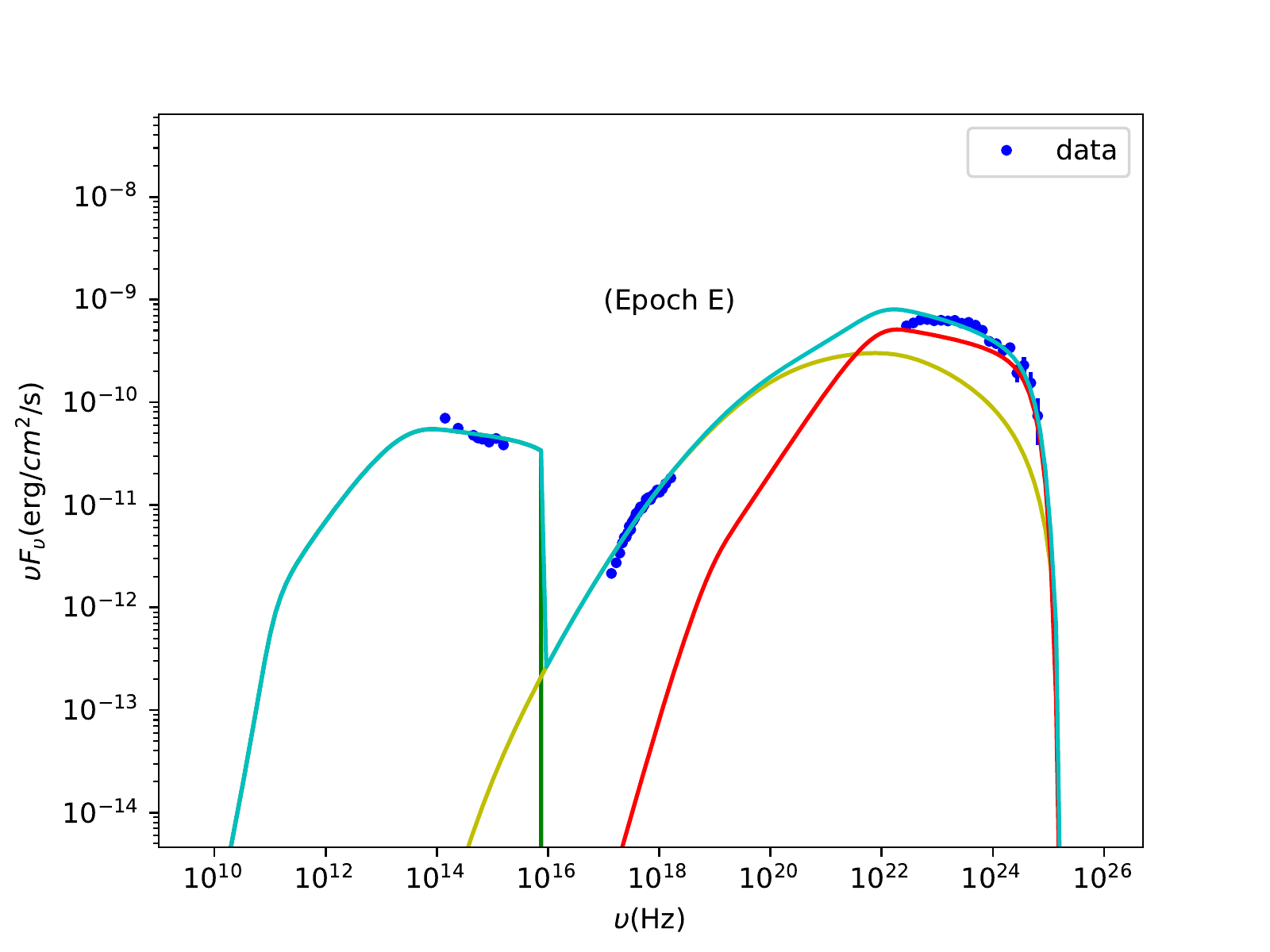}&
\includegraphics[width=95mm,height=75mm]{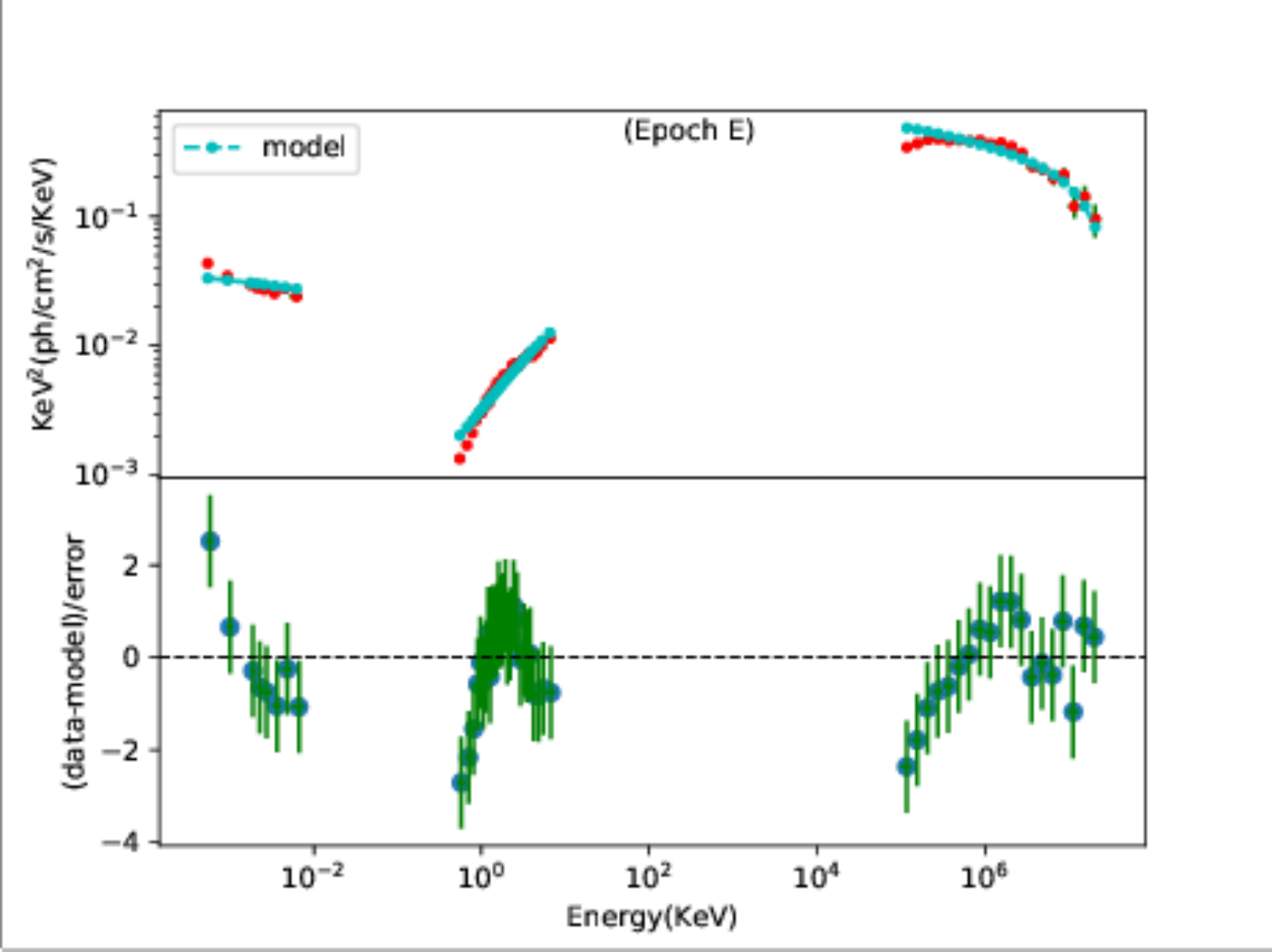}
\end{array}$
\end{center}
\caption{Model fits to the broad band SED during epochs D and E. The panels
have the same meanings as that of Fig. \ref{Fig12}}
\label{Fig13}
\end{figure*}

To investigate the difference in the flaring behavior of the source during 
various epochs, the SED fitting methodology was carried out in the following 
manner: First, the typical value of the source parameters governing the 
observed broadband emission from 3C\,454.3 was attained by fitting the 
quiescent state C. Limited information available through optical, X-ray 
and $\gamma$-ray observations did not let us to constrain all the parameters. 
The information that could be obtained from the observed SEDs were the high 
energy and low energy spectral indices, synchrotron flux at optical, SSC flux 
at X-ray and EC flux at $\gamma$-ray energies. Consistently we chose five 
parameters namely, $p$, $q$, $U_e$, $B$ and $\Gamma$ to be free and froze the 
rest of the parameters. Details of the parameters are given in Table\ref{Table:SED}. The values of $\gamma_{min}$, $\gamma_{max}$ and $\theta$ were chosen to be $40$, \textbf{$10^{4}$} and $1^{\circ}$ respectively. The 
other parameters have $R = 3\times10^{15}$, $T = 1000K$, $f = 0.90$. The adopted size of the $\gamma$-ray emitting region has been found from gravitational microlensing effect that ranges from $10^{14}-10^{15}$ cm \citep{2016A&A...586A.150V}. The resultant best fit parameters for the epoch C are given in 
Table \ref{Table:SED} and the model SED with the observed one is shown in Figure\ref{Fig12} and Fig. \ref{Fig13}.

The fitting procedure was repeated for the epochs A, B and D with the 
choice of free parameters similar to the case of epoch C. For epochs A and B, 
where optical and $\gamma$-ray flare are correlated, we found the main 
difference is seen in the enhancement of the bulk Lorentz factor and a 
marginal decrease in the magnetic field. For epoch A, the increase in bulk 
Lorentz factor is relatively less; however, this is also associated with an 
increase in electron energy density. On the other hand, for epoch D with an 
isolated orphan optical flare, we found the SED can be reproduced with an 
increase in the bulk Lorentz factor and magnetic field and decrease in electron energy density relative to the quiescent epoch C. For epoch E, where a major flare is observed in optical compared to the $\gamma$ ray,we found the SED cannot be reproduced satisfactorily with the parameters similar to epoch C rather it demands a large emission region size with low Lorentz factor. Hence the emission region during this epoch may be at large jet scale where the jet cross section is significantly larger.
During all the epochs, we also observed the variations 
in the high and low energy particle power-law indices and this can also 
manifest the flux variations observed at these energies. Our modeling
also shows that the observed broad band SED over all the epochs can be 
well described by the leptonic scenario (Fig. \ref{Fig12} and Fig. \ref{Fig13}).



\begin{table*}
\caption{Details of the PL and LP model fits for five epochs. Here the $\gamma$-ray flux value is in units of $10^{-6}$ph $cm^{-2}$ $s^{-1}$}
{\begin{tabular}{cccrrccrrrr}
     \hline
     &\multicolumn{4}{c}{PL} & \multicolumn{5}{c}{LP}&\\ 
     \cmidrule(lr){2-5} \cmidrule(lr){6-10}
     Epochs  & $\Gamma$ & Flux  & TS & $-$Log L  & $\alpha$ & $\beta$ & 
    Flux & TS & $-$Log L & TS$_{curve}$\\
     \hline
     A &-2.34$\pm$0.01&5.68$\pm$0.06& 87024.8 & 138177.8 & 1.91$\pm$0.03 & 0.15$\pm$ 0.01 &  4.93$\pm$0.07 &  79568.2 & 137962.7 & 430.1 \\
     B &-2.33$\pm$0.01&13.3$\pm$0.08&278585.0 &  80508.1 & 2.21$\pm$0.00 & 0.09$\pm$ 0.00 & 12.79$\pm$0.12 & 283811.0 &  80238.2 & 539.8 \\
     C &-2.42$\pm$0.03&0.57$\pm$0.02&  3422.6 & 151787.4 & 1.97$\pm$0.10 & 0.24$\pm$ 0.04 &  0.47$\pm$0.03 &   3392.8 & 151756.9 &  61.0 \\
     D &-2.25$\pm$0.02&1.03$\pm$0.03&  9890.6 & 142338.0 & 2.04$\pm$0.04 & 0.14$\pm$ 0.02 &  0.88$\pm$0.03 &   9768.6 & 142299.7 &  76.6 \\
     E &-2.14$\pm$0.00&4.42$\pm$0.00& 66161.9 & 146140.9 & 1.53$\pm$0.00 & 0.15$\pm$ 0.00 &  3.71$\pm$0.03 &  62145.3 & 145896.2 & 489.4 \\
     
     \hline     
     \end{tabular}}
     \label{Table:PL}
\end{table*}

\begin{table*}
\caption{Details of the SED fitting using XSPEC.The other adopted parameters for epochs A,B,C and D are $\gamma_{min}$ = 40, $\gamma_{max}$ = $10^{4}$, R = $3\times10^{15}$cm, T = 1000K, $\gamma_{b}$ = 1200, and f = 0.90. The parameters for epoch E are $\gamma_{min}$ = 40, $\gamma_{max}$ = $4.0\times10^{4}$, R = $9\times10^{16}$cm, T = 800K, $\gamma_{b}$ = 1500, and f = 0.80.}
{\begin{tabular}{lrrrrr}
     \hline
     Parameters& Epoch A & Epoch B & Epoch C & Epoch D & Epoch E \\
     \hline
     Bulk Lorentz Factor factor&11.79$\pm$0.82&16.63$\pm$0.96&7.30$\pm$0.30&11.88$\pm$1.07&5.00$\pm$0.71\\
     Low energy particle index&1.47$\pm$0.16&1.34$\pm$0.13&1.56$\pm$0.09&1.98$\pm$0.15&1.37$\pm$0.11\\
     High energy particle index&3.64$\pm$0.11&3.93$\pm$0.09&3.12$\pm$0.11&3.49$\pm$0.15&3.14$\pm$0.08\\
     Electron Energy Density (cm$^{-3})$ &5.15$\pm$1.08&2.05$\pm$0.36&3.62$\pm$0.32&1.15$\pm$0.24&0.04$\pm$0.01\\
     Magnetic Field (Gauss)&1.77$\pm$0.11&1.78$\pm$0.09&2.82$\pm$0.16&4.01$\pm$0.23&0.73$\pm$0.03\\
     Chi-square/dof&1.0~~~~~&1.0~~~~~&0.84~~~~~&1.30~~~~~&1.07~~~~~\\
     \hline     
\end{tabular}}
\label{Table:SED}
\end{table*}

\section{Results and discussion}\label{sec:results}
\subsection{$\gamma$-ray spectra}
\begin{figure}
\vbox{
\includegraphics[scale=0.6]{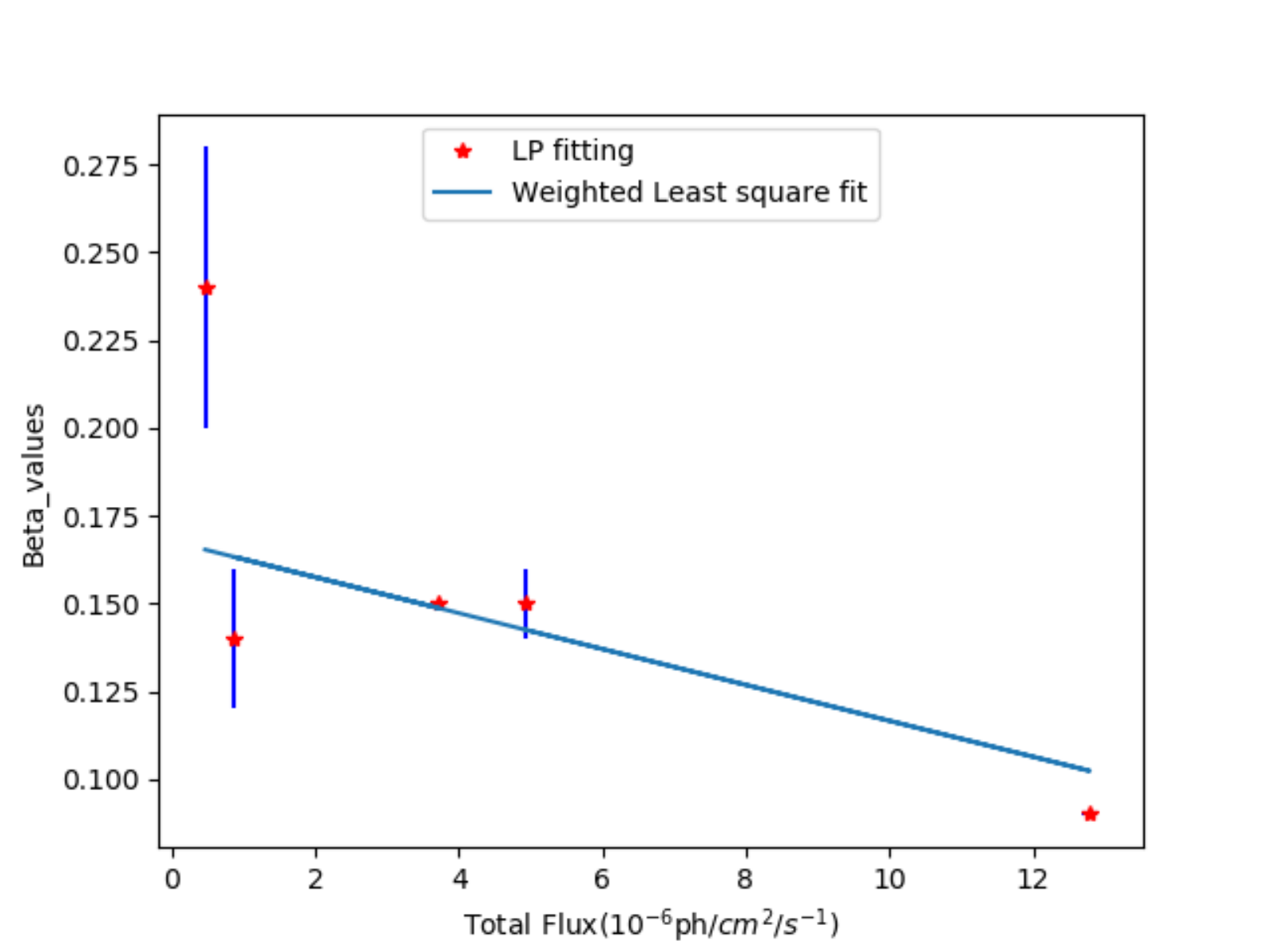}
\includegraphics[scale=0.6]{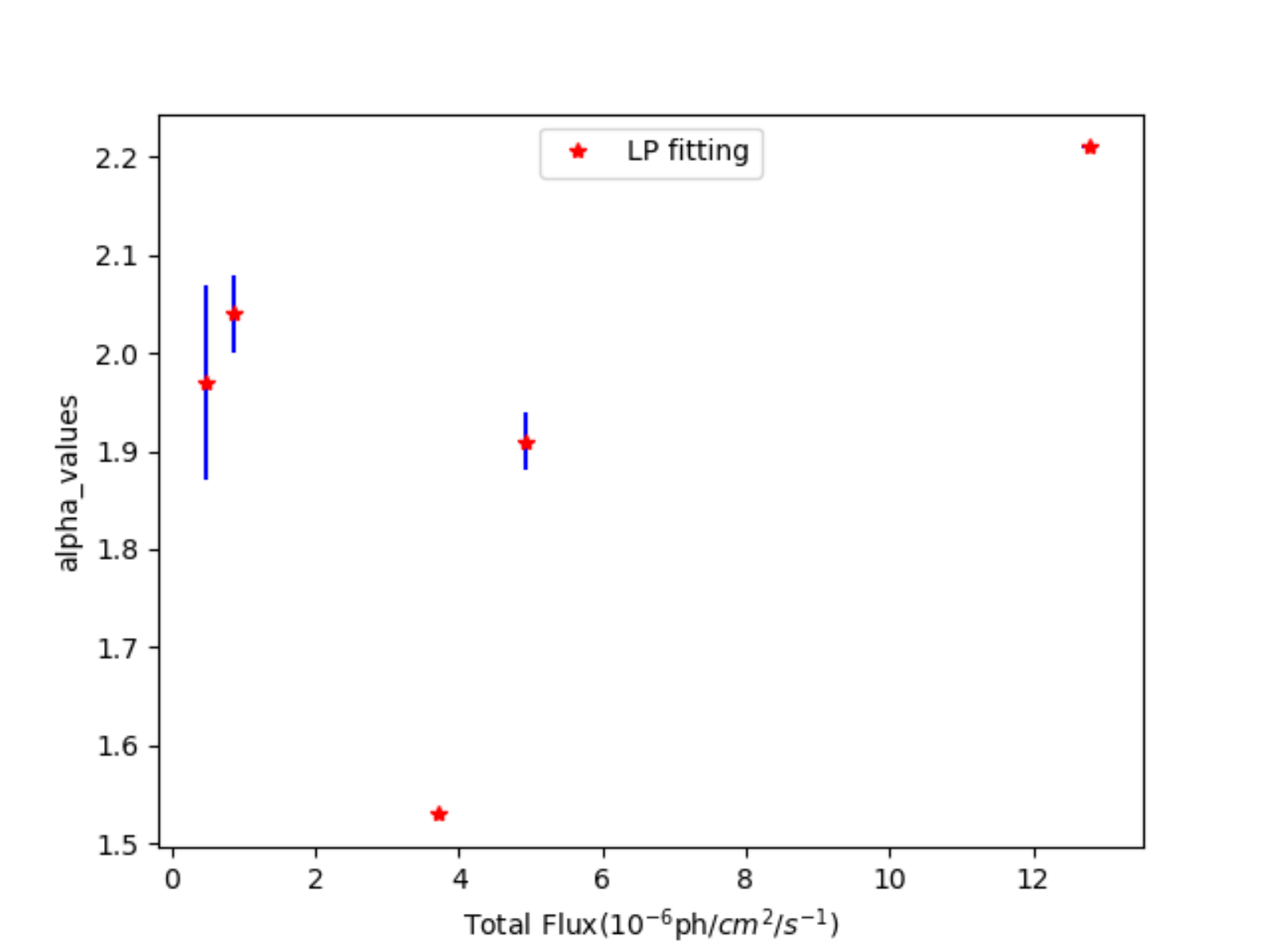}
}
\caption{Variations of the parameters $\alpha$ and $\beta$ with flux.}
\label{Fig14}
\end{figure}

The gamma-ray spectra of 3C 454.3 in all the five epochs considered here that
includes, both the active and quiescent phases are well described by the
LP model. Such curved $\gamma$-ray spectrum is generally seen in 
FSRQs \citep{2010ApJ...710.1271A}. In the third catalog of AGN detected
by  {\it Fermi} LAT, the spectrum of  3C 454.3
is well described by a power law with exponential cut-off model \citep{2015ApJ...810...14A}. Such an 
observed curved $\gamma$-ray spectrum could be  due to
the electrons giving rise to the emission having a curved energy 
distribution \citep{2015ApJ...809..174D}.
Alternatively, the curved $\gamma$-ray spectrum is a manifestation of the 
attenuation of high energy $\gamma$-rays through photon-photon 
pair production \citep{2016MNRAS.458..354C}. The parameter $\alpha$ in the 
LP model is a measure of the slope of the $\gamma$-ray spectrum with a 
small value of $\alpha$ indicating a harder $\gamma$-ray spectrum. The curvature parameter 
$\beta$ gives a measure of the presence of cut-off in the spectrum at high 
energies with a large value of $\beta$ indicating a sharper cut-off. 
Therefore, investigation of any changes in the values of $\alpha$ and $\beta$ parameters 
during the five epochs can point to changes in the $\gamma$-ray spectral 
shape. A change in the $\gamma$-ray spectral parameters during a
flaring state could point to change in the position of the $\gamma$-ray 
emitting region.  The variation of the $\alpha$ and 
$\beta$ against flux during the five epochs studied here is shown in Fig. 
\ref{Fig14}.  
No trend on the variation of $\alpha$ with flux was seen during the five
epochs considered here. However, a plot of $\beta$ with flux (Fig. \ref{Fig14})
shows a 
clear trend of a decrease in the curvature parameter with increasing flux. Linear least square fit to the data gives $\beta$ = (-0.005$\pm$0.009)$F_{V}$+(0.168$\pm$0.029) with a correlation coefficient of -0.77.
Clearly the value of $\beta$ is high and low at the lowest and the highest 
flux levels among the five epochs analyzed here.
Such a  trend is also seen by \cite{2015ApJ...810...14A} on
analysis of the FSRQs in the third catalog of AGN detected by Fermi (3LAC). The
changes in the $\gamma$-ray spectral shape during different brightness states
of 3C 454.3 could point to the emission site located at different regions 
during different brightness states, however, other explanations could not 
be ruled out.
Studies on the flares of 3C 454.3  
in December 2009 and November 2010, that falls in the epochs A and
B studied here concluded that  the $\gamma$-ray emission regions were located 
close to the central black hole \citep{2012arXiv1205.0520J,2013ApJ...779..100I}.

\subsection{Connection between optical and GeV flux variations}
\begin{figure}
\hspace*{-0.2cm}\includegraphics[scale=0.6]{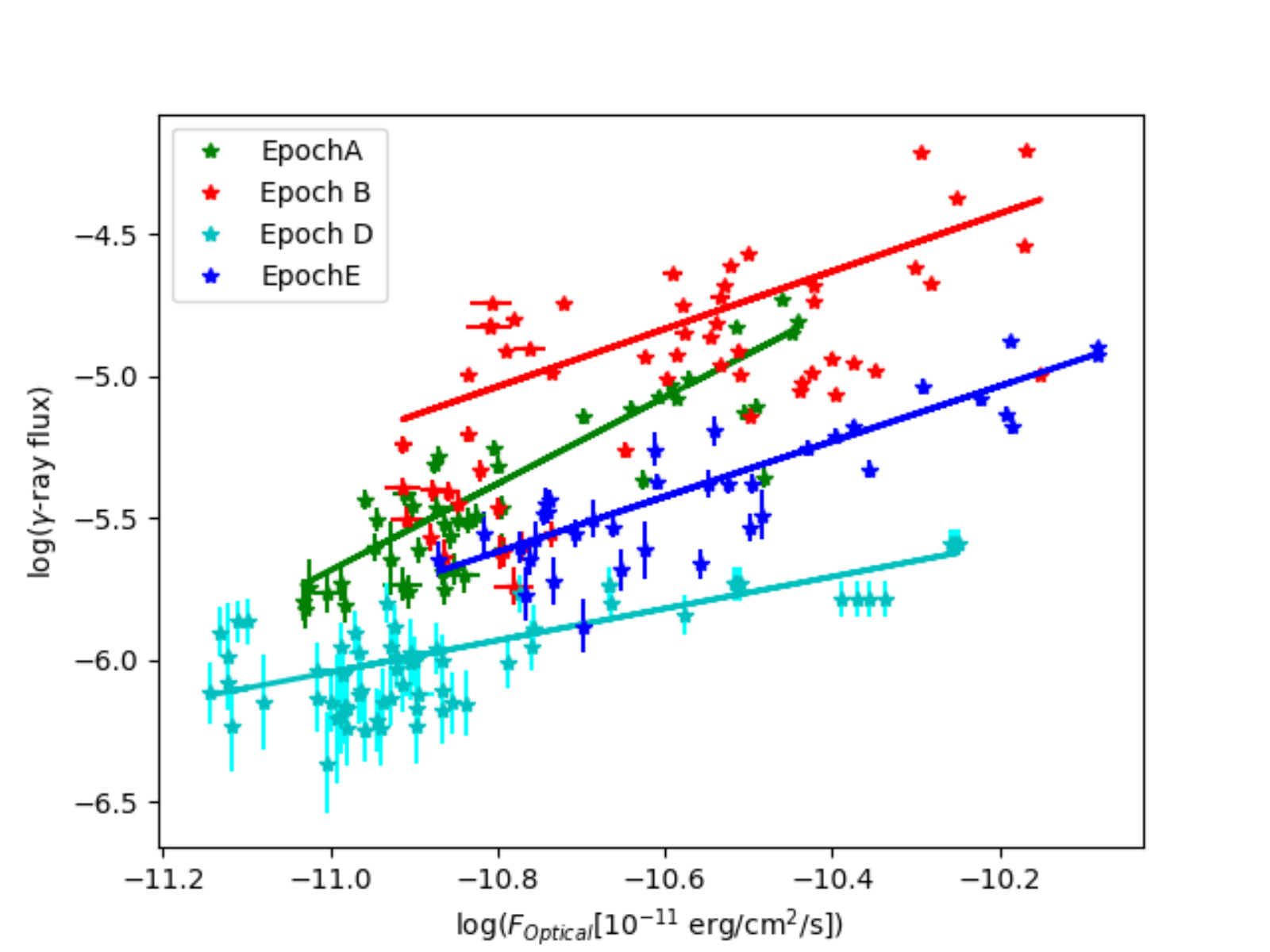}
\caption{Optical flux v/s $\gamma$-ray flux for epoch A,B,D and E.}
\label{Fig15}
\end{figure}

In a majority of the multi-wavelength monitoring observations of blazars, close
correlations between the flux variations in different bands were noticed 
\citep{2009ApJ...697L..81B}.  This was explained  on the co-spatiality of the 
emission regions emitting in different bands and the correlated optical and
GeV flux variations can be understood in the standard leptonic emission 
processes according to which the same relativistic electrons produce optical
and $\gamma$-ray emission via synchrotron and IC processes..  However, there are a handful of 
blazars where the emission in the optical and GeV $\gamma$-ray bands
are found not to be correlated \citep{2013ApJ...763L..11C,2014ApJ...797..137C,2013ApJ...779..174D,2015ApJ...804..111M}. During the 9 years of monitoring
data analyzed here we found four flaring epochs in the optical, namely
A,B, D and E. During epochs A and B, the optical flare is accompanied by a
$\gamma$-flare, while at the other two epochs D and E, though the optical
flares have amplitudes similar to that of epochs A and B, the $\gamma$-rays
during epochs D and E were either weak or undetected. This is clearly 
seen in Fig. \ref{Fig15}  where the logarithm of $\gamma$-ray flux is plotted 
against the logarithm of optical flux. The results of the linear least squares
fit carried out between the gamma-ray flux $F_{\gamma}$ and optical flux in the V-band $F_{V}$ during those four epochs yielded the 
following relations given in Equations 4-7  for epochs A, B, D and E
respectively. The results of the fit are given in 
Table \ref{Table:linear fits}. 

\begin{equation}
log F^{A}_{\gamma} = (1.528 \pm 0.112) log F^{A}_V + (11.128 \pm 1.190)
\end{equation}

\begin{equation}
log F^{B}_{\gamma} = (1.017 \pm 0.144) log F^{B}_V + (5.945 \pm 1.506)
\end{equation}

\begin{equation}
log F^{D}_{\gamma} = (0.561 \pm 0.048) log F^{D}_V + (0.132 \pm 0.510)
\end{equation}

\begin{equation}
log F^{E}_{\gamma} = (0.974 \pm 0.068) log F^{E}_V + (4.901 \pm 0.706)
\end{equation}

The above equations clearly indicate that when the source showed optical
flares during epochs D and E, the $\gamma$-ray emission is weak. Thus during these two epochs there is a clear case of 
optical flares with weak/no-corresponding $\gamma$-ray counterparts. Also during epoch B, prior to the large optical flare with a counterpart in the gamma-ray band, a short duration and intense optical flare was found around MJD 55510, without a $\gamma$-ray counterpart also only noticed by \cite{2011ApJ...736L..38V}. According to \cite{2011ApJ...736L..38V} such a lack of a simultaneous $\gamma$-ray at MJD 55510, could be due to either enhancement of the magnetic field, or attenuation by $\gamma$-$\gamma$ production or lack of external seed photons. However, based on arguments from modeling \cite{2011ApJ...736L..38V} indicate that the complex behaviour seen during epoch B could be due to changes in the external photon field. However, according to \cite{2014ApJ...793...98V} the anomalous flux variability patterns between optical and $\gamma$-ray can be due to inverse compton scattering or process happening as the jet collides onto mirror cloud situated at parsec scales. Thus
the variability shown by 3C 454.3 in different energy bands is complex.
To have an insight into this anomalous variability behavior we fitted
the broad band SED of the source in all the five epochs using simple one
zone leptonic emission models. During epochs A and B, where the optical and
$\gamma$-ray flux variations are correlated, there is enhancement in the 
bulk Lorentz factor relative to the quiescent epoch C. During epoch D,
we found an enhancement of the magnetic field related to the quiescent state C, which could explain the high optical 
flare accompanied by a very weak $\gamma$-ray flare. Such a change in magnetic
field could also produce enhanced optical polarization and X-ray flux.
But, the non-availability of optical polarization and X-ray flux measurements
during epoch D, preclude us to make a firm conclusion on the enhancement
of the magnetic field as the cause for the occurrence of optical flare with weak
$\gamma$-ray flare during epoch D, however, is the most favorable scenario. 
In epoch E,where there is an optical flare with a weak $\gamma$-ray counterpart
our SED modeling indicates decrease in electron energy density, magnetic field and bulk Lorentz factor and also the emission region could be located at a region farther than the emission region of other epochs. We therefore conclude that 
the observations of optical flare with weak/no corresponding $\gamma$-ray flare 
during epochs D and E, could be due to one or a combination of parameters such as 
the bulk Lorentz factor, magnetic field and electron energy density or due to 
changes in the locations of the $\gamma$-ray emitting regions. 3C 454.3 is fourth
blazar known to have shown the anomalous variability behavior of
optical flares with no $\gamma$-counterparts. The other sources where such
behavior were noticed are PKS 0208-512 \citep{2013ApJ...763L..11C}, 
S4 1849+67 \citep{2014ApJ...797..137C} and 3C 279 \citep{2018MNRAS.479.2037P}.  
A possible cause for optical flux variations without $\gamma$-ray counterparts
could be attributed to hadronic processes \citep{2001APh....15..121M}, however
based on our SED analysis, we conclude that the leptonic model is also capable of explaining
the emission from 3C 454.3 during all the epochs.

\subsection{Correlation between optical flux and polarization}
During Epoch B, there is one short term optical flare at around 55510 MJD. During this 
period there is no $\gamma$-ray flare. This intense short duration optical flare without any $\gamma$-ray counterpart was also reported by \cite{2011ApJ...736L..38V}. Such an increased optical flare without any $\gamma$-ray counterpart can be due to magnetic field enhancement. In this scenario an increase in magnetic field will lead to increased optical flare (from synchrotron process) and no increased $\gamma$-ray emission as the $\gamma$-ray emission from inverse compton process is independent of magnetic field \citep{2011ApJ...736L..38V,2013ApJ...771L..25C}. Alternatively, the lack of a $\gamma$-ray flare coinciding with an optical flare at around 55510 MJD could be due to the attenuation of $\gamma$-ray via pair production or lack of external photon field \cite{2011ApJ...736L..38V}. Polarization observations can play a key role in arriving at a possible scenario for the anomalous flare seen at MJD 55510. During this time sparse 
polarization observations were available to make any analysis on the 
correlation between flux and polarization variations possible. Degree of 
polarization seems to be positively correlated with the optical flux change 
(Fig. \ref{Fig4}) with the increase in the degree of polarization coinciding with the 
optical flare. Near simultaneous polarization observations were also available 
for epoch E. 
During this epoch, we have sufficient photometric observations
to study the correlation between the degree of polarization and flux changes.
The degree of polarization
is found to be anti-correlated to the flux changes in the optical V-band, both
during the rising phase as well as the declining phase of the optical flare 
evident in Fig.\ref{Fig8}. Such, anti-correlation between degree of optical polarization 
and total flux is known before in the BL Lacertae object BL Lac 
\citep{2014ApJ...781L...4G} and the FSRQ 3C 454.3 \citep{2017MNRAS.472..788G}, 
Such anti-correlation between flux and degree
of polarization  could be explained in a two component model, consisting
of a slowing varying component and a variable components with different
polarization directions. 

\begin{table}
\caption{Results of the linear least squares fit to the 
optical and $\gamma$-ray flux measurements, during epochs A,B,D and E}
{\begin{tabular}{ccrc}
     \hline
     Epoch & Slope & Intercept & Correlation coefficient \\
     \hline
     A & 1.528$\pm$0.112 & 11.128$\pm$1.190 & 0.880 \\
     B & 1.017$\pm$0.144 &  5.945$\pm$1.506 & 0.680 \\
     D & 0.561$\pm$0.048 &  0.132$\pm$0.510 & 0.742 \\
     E & 0.974$\pm$0.068 &  4.901$\pm$0.706 & 0.860 \\

     \hline
     \end{tabular}}
     \label{Table:linear fits}
\end{table}

\section{Summary}\label{sec:summary}
We present our analysis of  the multiband light curves of the FSRQ 3C 454.3 
that include $\gamma$-rays, X-rays, UV, optical and IR  
spanning about 9 years from 2008 August to 2017 February. The 
results are summarized below:
\begin{enumerate}
\item Between the period 2008 August to 2017 February, 3C 454.3 showed large amplitude optical/IR flares during four epochs identified in this work as A, B, D and E. During epoch A, the optical flare has a counterpart in the $\gamma$-ray region. Cross correlation analysis indicated that the optical and $\gamma$-ray flux variations are closely correlated with a lag of 2.5$^{+1.5}_{-1.4}$ days with the optical lagging the $\gamma$-ray emission, pointing to difference in their emission regions. During epoch B, too, the optical flare has a corresponding $\gamma$-ray flare. From cross-correlation analysis we found that the flux variations in both the optical and $\gamma$-band are correlated with almost zero lag, pointing to co-spatiality of both the emission regions. During epochs D and E, though optical flare has similar magnitude to that of the flare at epochs A and B, the source is weak in the $\gamma$-ray band. Our analysis thus points to the detection of optical flare with no $\gamma$-ray counterpart in 3C 454.3. The only other sources where such behaviour were observed are PKS 0208$-$512 \citep{2013ApJ...763L..11C}, S4 1849+67 \citep{2014ApJ...797..137C} and 3C 279 \citep{2018MNRAS.479.2037P}.
\item Broad band SED modeling, using one zone leptonic emission model was carried out on the four flaring epochs A,B,D and E along with a quiescent epoch C for comparison. Relative to the quiescent state C, during the active state A and B, there is enhancement of the bulk Lorentz factor, which could explain the correlated optical and $\gamma$-ray flux variations. The observations of optical flare with weak/no $\gamma$-ray counterpart during epochs D and E, could be due to a combination of parameters such as the bulk Lorentz factor, magnetic field and electron energy density or changes in the location of the $\gamma$-ray emitting region.
\item The available polarization observations during the period analysed here showed complex correlation to the optical flux changes. During a short term optical flare (apart from the main flare) during epoch B, the degree of optical polarization was found to be correlated to the optical flux changes. However, during the flare at epoch E, the degree of optical polarization is anti-correlated to the optical V-band brightness both during the rising and falling part of the flare.
\item The source showed a complex colour (V-J)-magnitude (V) variability. During the epochs A and B, when optical and $\gamma$- ray variations are correlated, the source showed a RWB behaviour. For epoch D, the source showed a RWB trend for V-band brightness fainter than 15 mag, while for V-band magnitude brighter than 15 mag, it showed a BWB behaviour. For epoch E, when there is an optical flare with no $\gamma$-ray counterpart, the source showed a BWB trend.
\item The $\gamma$-ray spectra during all the five epochs were well described by a LP model. The curvature parameter $\beta$ that provides an indication of the cut-off present in the spectrum is found to be lowest at the highest flux level and highest at the lowest flux level among the five epochs analyzed here.
\end{enumerate}

\section*{Acknowledgements}
We thank the anonymous referee for his/her critical comments that helped to improve the manuscript. This research has made use of data, software and web tools of High Energy Astrophysics Science Archive Research center (HEASARC), maintained by NASA’s Goddard Space Flight Center. SMARTS observations of LAT-monitored blazars are supported by Yale University and Fermi  GI  grant  NNX  12AP15G.  J.C.I. has received support from NASA-Harriet Jenkins Pre-doctoral Fellowship Program, NSF Graduate Research Fellowship Program (DGE-0644492), and the National Research Council’s Ford Foundation Disserta-
tion Fellowship. C.D.B., M.M.B, and the SMARTS 1.3 m and
1.5 m observing queue also receive support from NSF grant AST-0707627. We are grateful for photometry and polarimetry from Paul Smith’s monitoring program at the Steward Observatory, which is supported by Fermi Guest Investigator grants NNX08AW56G, NNX09AU10G, and NNX12AO93G.


\bibliographystyle{mnras}
\bibliography{ref.bib}

\begin{thebibliography}{}
\makeatletter
\relax
\def\mn@urlcharsother{\let\do\@makeother \do\$\do\&\do\#\do\^\do\_\do\%\do\~}
\def\mn@doi{\begingroup\mn@urlcharsother \@ifnextchar [ {\mn@doi@}
  {\mn@doi@[]}}
\def\mn@doi@[#1]#2{\def\@tempa{#1}\ifx\@tempa\@empty \href
  {http://dx.doi.org/#2} {doi:#2}\else \href {http://dx.doi.org/#2} {#1}\fi
  \endgroup}
\def\mn@eprint#1#2{\mn@eprint@#1:#2::\@nil}
\def\mn@eprint@arXiv#1{\href {http://arxiv.org/abs/#1} {{\tt arXiv:#1}}}
\def\mn@eprint@dblp#1{\href {http://dblp.uni-trier.de/rec/bibtex/#1.xml}
  {dblp:#1}}
\def\mn@eprint@#1:#2:#3:#4\@nil{\def\@tempa {#1}\def\@tempb {#2}\def\@tempc
  {#3}\ifx \@tempc \@empty \let \@tempc \@tempb \let \@tempb \@tempa \fi \ifx
  \@tempb \@empty \def\@tempb {arXiv}\fi \@ifundefined
  {mn@eprint@\@tempb}{\@tempb:\@tempc}{\expandafter \expandafter \csname
  mn@eprint@\@tempb\endcsname \expandafter{\@tempc}}}

\bibitem[\protect\citeauthoryear{{Abdo} et~al.,}{{Abdo}
  et~al.}{2010a}]{2010ApJ...710.1271A}
{Abdo} A.~A.,  et~al., 2010a, \mn@doi [\apj] {10.1088/0004-637X/710/2/1271},
  \href {http://adsabs.harvard.edu/abs/2010ApJ...710.1271A} {710, 1271}

\bibitem[\protect\citeauthoryear{{Abdo} et~al.,}{{Abdo}
  et~al.}{2010b}]{2010ApJ...716...30A}
{Abdo} A.~A.,  et~al., 2010b, \mn@doi [\apj] {10.1088/0004-637X/716/1/30},
  \href {http://adsabs.harvard.edu/abs/2010ApJ...716...30A} {716, 30}

\bibitem[\protect\citeauthoryear{{Abdo} et~al.,}{{Abdo}
  et~al.}{2011}]{2011ApJ...733L..26A}
{Abdo} A.~A.,  et~al., 2011, \mn@doi [\apjl] {10.1088/2041-8205/733/2/L26},
  \href {http://adsabs.harvard.edu/abs/2011ApJ...733L..26A} {733, L26}

\bibitem[\protect\citeauthoryear{{Acero} et~al.,}{{Acero}
  et~al.}{2015}]{2015ApJS..218...23A}
{Acero} F.,  et~al., 2015, \mn@doi [\apjs] {10.1088/0067-0049/218/2/23}, \href
  {http://adsabs.harvard.edu/abs/2015ApJS..218...23A} {218, 23}

\bibitem[\protect\citeauthoryear{{Ackermann} et~al.,}{{Ackermann}
  et~al.}{2010}]{2010ApJ...721.1383A}
{Ackermann} M.,  et~al., 2010, \mn@doi [\apj] {10.1088/0004-637X/721/2/1383},
  \href {http://adsabs.harvard.edu/abs/2010ApJ...721.1383A} {721, 1383}

\bibitem[\protect\citeauthoryear{{Ackermann} et~al.,}{{Ackermann}
  et~al.}{2015}]{2015ApJ...810...14A}
{Ackermann} M.,  et~al., 2015, \mn@doi [\apj] {10.1088/0004-637X/810/1/14},
  \href {http://adsabs.harvard.edu/abs/2015ApJ...810...14A} {810, 14}

\bibitem[\protect\citeauthoryear{{Antonucci}}{{Antonucci}}{1993}]{1993ARA&A..31..473A}
{Antonucci} R.,  1993, \mn@doi [\araa] {10.1146/annurev.aa.31.090193.002353},
  \href {http://cdsads.u-strasbg.fr/abs/1993ARA%26A..31..473A} {31, 473}

\bibitem[\protect\citeauthoryear{{Arnaud}}{{Arnaud}}{1996}]{1996ASPC..101...17A}
{Arnaud} K.~A.,  1996, in {Jacoby} G.~H.,  {Barnes} J.,  eds,  Astronomical
  Society of the Pacific Conference Series Vol. 101, Astronomical Data Analysis
  Software and Systems V. p.~17

\bibitem[\protect\citeauthoryear{{Atwood} et~al.,}{{Atwood}
  et~al.}{2009}]{2009ApJ...697.1071A}
{Atwood} W.~B.,  et~al., 2009, \mn@doi [\apj] {10.1088/0004-637X/697/2/1071},
  \href {http://adsabs.harvard.edu/abs/2009ApJ...697.1071A} {697, 1071}

\bibitem[\protect\citeauthoryear{{Begelman} et~al.,}{{Begelman}
  et~al.}{1987}]{1987ApJ...322..650B}
{Begelman} M.~C.,  et~al., 1987, \mn@doi [\apj] {10.1086/165760}, \href
  {http://adsabs.harvard.edu/abs/1987ApJ...322..650B} {322, 650}

\bibitem[\protect\citeauthoryear{{B{\l}a{\.z}ejowski}, {Sikora}, {Moderski}  \&
  {Madejski}}{{B{\l}a{\.z}ejowski} et~al.}{2000}]{2000ApJ...545..107B}
{B{\l}a{\.z}ejowski} M.,  {Sikora} M.,  {Moderski} R.,   {Madejski} G.~M.,
  2000, \mn@doi [\apj] {10.1086/317791}, \href
  {http://adsabs.harvard.edu/abs/2000ApJ...545..107B} {545, 107}

\bibitem[\protect\citeauthoryear{{Boettcher}, {Mause}  \&
  {Schlickeiser}}{{Boettcher} et~al.}{1997}]{1997A&A...324..395B}
{Boettcher} M.,  {Mause} H.,   {Schlickeiser} R.,  1997, \aap, \href
  {http://adsabs.harvard.edu/abs/1997A%26A...324..395B} {324, 395}

\bibitem[\protect\citeauthoryear{{Bonning} et~al.,}{{Bonning}
  et~al.}{2009}]{2009ApJ...697L..81B}
{Bonning} E.~W.,  et~al., 2009, \mn@doi [\apjl] {10.1088/0004-637X/697/2/L81},
  \href {http://adsabs.harvard.edu/abs/2009ApJ...697L..81B} {697, L81}

\bibitem[\protect\citeauthoryear{{Bonning} et~al.,}{{Bonning}
  et~al.}{2012}]{2012ApJ...756...13B}
{Bonning} E.,  et~al., 2012, \mn@doi [\apj] {10.1088/0004-637X/756/1/13}, \href
  {http://adsabs.harvard.edu/abs/2012ApJ...756...13B} {756, 13}

\bibitem[\protect\citeauthoryear{{B{\"o}ttcher}}{{B{\"o}ttcher}}{2007}]{2007Ap&SS.309...95B}
{B{\"o}ttcher} M.,  2007, \mn@doi [\apss] {10.1007/s10509-007-9404-0}, \href
  {http://adsabs.harvard.edu/abs/2007Ap%26SS.309...95B} {309, 95}

\bibitem[\protect\citeauthoryear{{B{\"o}ttcher}, {Reimer}, {Sweeney}  \&
  {Prakash}}{{B{\"o}ttcher} et~al.}{2013}]{2013ApJ...768...54B}
{B{\"o}ttcher} M.,  {Reimer} A.,  {Sweeney} K.,   {Prakash} A.,  2013, \mn@doi
  [\apj] {10.1088/0004-637X/768/1/54}, \href
  {http://adsabs.harvard.edu/abs/2013ApJ...768...54B} {768, 54}

\bibitem[\protect\citeauthoryear{{Breeveld}, {Landsman}, {Holland}, {Roming},
  {Kuin}  \& {Page}}{{Breeveld} et~al.}{2011}]{2011AIPC.1358..373B}
{Breeveld} A.~A.,  {Landsman} W.,  {Holland} S.~T.,  {Roming} P.,  {Kuin}
  N.~P.~M.,   {Page} M.~J.,  2011, in {McEnery} J.~E.,  {Racusin} J.~L.,
  {Gehrels} N.,  eds,  American Institute of Physics Conference Series Vol.
  1358, American Institute of Physics Conference Series. pp 373--376
  (\mn@eprint {arXiv} {1102.4717}), \mn@doi{10.1063/1.3621807}

\bibitem[\protect\citeauthoryear{{Burrows} et~al.,}{{Burrows}
  et~al.}{2005}]{2005SSRv..120..165B}
{Burrows} D.~N.,  et~al., 2005, \mn@doi [\ssr] {10.1007/s11214-005-5097-2},
  \href {http://adsabs.harvard.edu/abs/2005SSRv..120..165B} {120, 165}

\bibitem[\protect\citeauthoryear{{Carnerero} et~al.,}{{Carnerero}
  et~al.}{2015}]{2015MNRAS.450.2677C}
{Carnerero} M.~I.,  et~al., 2015, \mn@doi [\mnras] {10.1093/mnras/stv823},
  \href {http://adsabs.harvard.edu/abs/2015MNRAS.450.2677C} {450, 2677}

\bibitem[\protect\citeauthoryear{{Chatterjee} et~al.,}{{Chatterjee}
  et~al.}{2012}]{2012ApJ...749..191C}
{Chatterjee} R.,  et~al., 2012, \mn@doi [\apj] {10.1088/0004-637X/749/2/191},
  \href {http://adsabs.harvard.edu/abs/2012ApJ...749..191C} {749, 191}

\bibitem[\protect\citeauthoryear{{Chatterjee} et~al.,}{{Chatterjee}
  et~al.}{2013a}]{2013ApJ...763L..11C}
{Chatterjee} R.,  et~al., 2013a, \mn@doi [\apjl] {10.1088/2041-8205/763/1/L11},
  \href {http://adsabs.harvard.edu/abs/2013ApJ...763L..11C} {763, L11}

\bibitem[\protect\citeauthoryear{{Chatterjee}, {Nalewajko}  \&
  {Myers}}{{Chatterjee} et~al.}{2013b}]{2013ApJ...771L..25C}
{Chatterjee} R.,  {Nalewajko} K.,   {Myers} A.~D.,  2013b, \mn@doi [\apjl]
  {10.1088/2041-8205/771/2/L25}, \href
  {http://adsabs.harvard.edu/abs/2013ApJ...771L..25C} {771, L25}

\bibitem[\protect\citeauthoryear{{Cohen}, {Romani}, {Filippenko}, {Cenko},
  {Lott}, {Zheng}  \& {Li}}{{Cohen} et~al.}{2014}]{2014ApJ...797..137C}
{Cohen} D.~P.,  {Romani} R.~W.,  {Filippenko} A.~V.,  {Cenko} S.~B.,  {Lott}
  B.,  {Zheng} W.,   {Li} W.,  2014, \mn@doi [\apj]
  {10.1088/0004-637X/797/2/137}, \href
  {http://adsabs.harvard.edu/abs/2014ApJ...797..137C} {797, 137}

\bibitem[\protect\citeauthoryear{{Coogan}, {Brown}  \& {Chadwick}}{{Coogan}
  et~al.}{2016}]{2016MNRAS.458..354C}
{Coogan} R.~T.,  {Brown} A.~M.,   {Chadwick} P.~M.,  2016, \mn@doi [\mnras]
  {10.1093/mnras/stw199}, \href
  {http://adsabs.harvard.edu/abs/2016MNRAS.458..354C} {458, 354}

\bibitem[\protect\citeauthoryear{{Dermer} \& {Schlickeiser}}{{Dermer} \&
  {Schlickeiser}}{1993}]{1993ApJ...416..458D}
{Dermer} C.~D.,  {Schlickeiser} R.,  1993, \mn@doi [\apj] {10.1086/173251},
  \href {http://adsabs.harvard.edu/abs/1993ApJ...416..458D} {416, 458}

\bibitem[\protect\citeauthoryear{{Dermer}, {Yan}, {Zhang}, {Finke}  \&
  {Lott}}{{Dermer} et~al.}{2015}]{2015ApJ...809..174D}
{Dermer} C.~D.,  {Yan} D.,  {Zhang} L.,  {Finke} J.~D.,   {Lott} B.,  2015,
  \mn@doi [\apj] {10.1088/0004-637X/809/2/174}, \href
  {http://adsabs.harvard.edu/abs/2015ApJ...809..174D} {809, 174}

\bibitem[\protect\citeauthoryear{{Diltz} \& {B{\"o}ttcher}}{{Diltz} \&
  {B{\"o}ttcher}}{2016}]{2016ApJ...826...54D}
{Diltz} C.,  {B{\"o}ttcher} M.,  2016, \mn@doi [\apj]
  {10.3847/0004-637X/826/1/54}, \href
  {http://adsabs.harvard.edu/abs/2016ApJ...826...54D} {826, 54}

\bibitem[\protect\citeauthoryear{{Dutka} et~al.,}{{Dutka}
  et~al.}{2013}]{2013ApJ...779..174D}
{Dutka} M.~S.,  et~al., 2013, \mn@doi [\apj] {10.1088/0004-637X/779/2/174},
  \href {http://adsabs.harvard.edu/abs/2013ApJ...779..174D} {779, 174}

\bibitem[\protect\citeauthoryear{{Edelson} \& {Krolik}}{{Edelson} \&
  {Krolik}}{1988}]{1988ApJ...333..646E}
{Edelson} R.~A.,  {Krolik} J.~H.,  1988, \mn@doi [\apj] {10.1086/166773}, \href
  {http://cdsads.u-strasbg.fr/abs/1988ApJ...333..646E} {333, 646}

\bibitem[\protect\citeauthoryear{{Fossati}, {Maraschi}, {Celotti}, {Comastri}
  \& {Ghisellini}}{{Fossati} et~al.}{1998}]{1998MNRAS.299..433F}
{Fossati} G.,  {Maraschi} L.,  {Celotti} A.,  {Comastri} A.,   {Ghisellini} G.,
   1998, \mn@doi [\mnras] {10.1046/j.1365-8711.1998.01828.x}, \href
  {http://adsabs.harvard.edu/abs/1998MNRAS.299..433F} {299, 433}

\bibitem[\protect\citeauthoryear{{Gaskell} \& {Peterson}}{{Gaskell} \&
  {Peterson}}{1987}]{1987ApJS...65....1G}
{Gaskell} C.~M.,  {Peterson} B.~M.,  1987, \mn@doi [\apjs] {10.1086/191216},
  \href {http://adsabs.harvard.edu/abs/1987ApJS...65....1G} {65, 1}

\bibitem[\protect\citeauthoryear{{Gaskell} \& {Sparke}}{{Gaskell} \&
  {Sparke}}{1986}]{1986ApJ...305..175G}
{Gaskell} C.~M.,  {Sparke} L.~S.,  1986, \mn@doi [\apj] {10.1086/164238}, \href
  {http://cdsads.u-strasbg.fr/abs/1986ApJ...305..175G} {305, 175}

\bibitem[\protect\citeauthoryear{{Gaur}, {Gupta}  \& {Wiita}}{{Gaur}
  et~al.}{2012}]{2012AJ....143...23G}
{Gaur} H.,  {Gupta} A.~C.,   {Wiita} P.~J.,  2012, \mn@doi [\aj]
  {10.1088/0004-6256/143/1/23}, \href
  {http://adsabs.harvard.edu/abs/2012AJ....143...23G} {143, 23}

\bibitem[\protect\citeauthoryear{{Gaur}, {Gupta}, {Wiita}, {Uemura}, {Itoh}  \&
  {Sasada}}{{Gaur} et~al.}{2014}]{2014ApJ...781L...4G}
{Gaur} H.,  {Gupta} A.~C.,  {Wiita} P.~J.,  {Uemura} M.,  {Itoh} R.,   {Sasada}
  M.,  2014, \mn@doi [\apjl] {10.1088/2041-8205/781/1/L4}, \href
  {http://adsabs.harvard.edu/abs/2014ApJ...781L...4G} {781, L4}

\bibitem[\protect\citeauthoryear{{Gehrels} et~al.,}{{Gehrels}
  et~al.}{2004}]{2004ApJ...611.1005G}
{Gehrels} N.,  et~al., 2004, \mn@doi [\apj] {10.1086/422091}, \href
  {http://adsabs.harvard.edu/abs/2004ApJ...611.1005G} {611, 1005}

\bibitem[\protect\citeauthoryear{{Ghisellini} \& {Madau}}{{Ghisellini} \&
  {Madau}}{1996}]{1996MNRAS.280...67G}
{Ghisellini} G.,  {Madau} P.,  1996, \mn@doi [\mnras] {10.1093/mnras/280.1.67},
  \href {http://adsabs.harvard.edu/abs/1996MNRAS.280...67G} {280, 67}

\bibitem[\protect\citeauthoryear{{Ghisellini} \& {Maraschi}}{{Ghisellini} \&
  {Maraschi}}{1989}]{1989ApJ...340..181G}
{Ghisellini} G.,  {Maraschi} L.,  1989, \mn@doi [\apj] {10.1086/167383}, \href
  {http://adsabs.harvard.edu/abs/1989ApJ...340..181G} {340, 181}

\bibitem[\protect\citeauthoryear{{Ghisellini} \& {Tavecchio}}{{Ghisellini} \&
  {Tavecchio}}{2008}]{2008MNRAS.387.1669G}
{Ghisellini} G.,  {Tavecchio} F.,  2008, \mn@doi [\mnras]
  {10.1111/j.1365-2966.2008.13360.x}, \href
  {http://adsabs.harvard.edu/abs/2008MNRAS.387.1669G} {387, 1669}

\bibitem[\protect\citeauthoryear{{Gu}, {Lee}, {Pak}, {Yim}  \& {Fletcher}}{{Gu}
  et~al.}{2006}]{2006A&A...450...39G}
{Gu} M.~F.,  {Lee} C.-U.,  {Pak} S.,  {Yim} H.~S.,   {Fletcher} A.~B.,  2006,
  \mn@doi [\aap] {10.1051/0004-6361:20054271}, \href
  {http://adsabs.harvard.edu/abs/2006A%26A...450...39G} {450, 39}

\bibitem[\protect\citeauthoryear{{Gupta} et~al.,}{{Gupta}
  et~al.}{2017}]{2017MNRAS.472..788G}
{Gupta} A.~C.,  et~al., 2017, \mn@doi [\mnras] {10.1093/mnras/stx2072}, \href
  {http://adsabs.harvard.edu/abs/2017MNRAS.472..788G} {472, 788}

\bibitem[\protect\citeauthoryear{{Hartman} et~al.,}{{Hartman}
  et~al.}{1993}]{1993ApJ...407L..41H}
{Hartman} R.~C.,  et~al., 1993, \mn@doi [\apjl] {10.1086/186801}, \href
  {http://adsabs.harvard.edu/abs/1993ApJ...407L..41H} {407, L41}

\bibitem[\protect\citeauthoryear{{Hartman} et~al.,}{{Hartman}
  et~al.}{1999}]{1999ApJS..123...79H}
{Hartman} R.~C.,  et~al., 1999, \mn@doi [\apjs] {10.1086/313231}, \href
  {http://adsabs.harvard.edu/abs/1999ApJS..123...79H} {123, 79}

\bibitem[\protect\citeauthoryear{{Isler} et~al.,}{{Isler}
  et~al.}{2013}]{2013ApJ...779..100I}
{Isler} J.~C.,  et~al., 2013, \mn@doi [\apj] {10.1088/0004-637X/779/2/100},
  \href {http://adsabs.harvard.edu/abs/2013ApJ...779..100I} {779, 100}

\bibitem[\protect\citeauthoryear{{Jorstad} et~al.,}{{Jorstad}
  et~al.}{2012}]{2012arXiv1205.0520J}
{Jorstad} S.~G.,  et~al., 2012, preprint, \href
  {http://adsabs.harvard.edu/abs/2012arXiv1205.0520J} {} (\mn@eprint {arXiv}
  {1205.0520})

\bibitem[\protect\citeauthoryear{{Kalberla}, {Burton}, {Hartmann}, {Arnal},
  {Bajaja}, {Morras}  \& {P{\"o}ppel}}{{Kalberla}
  et~al.}{2005}]{2005A&A...440..775K}
{Kalberla} P.~M.~W.,  {Burton} W.~B.,  {Hartmann} D.,  {Arnal} E.~M.,  {Bajaja}
  E.,  {Morras} R.,   {P{\"o}ppel} W.~G.~L.,  2005, \mn@doi [\aap]
  {10.1051/0004-6361:20041864}, \href
  {http://adsabs.harvard.edu/abs/2005A%26A...440..775K} {440, 775}

\bibitem[\protect\citeauthoryear{{Konigl}}{{Konigl}}{1981}]{1981ApJ...243..700K}
{Konigl} A.,  1981, \mn@doi [\apj] {10.1086/158638}, \href
  {http://adsabs.harvard.edu/abs/1981ApJ...243..700K} {243, 700}

\bibitem[\protect\citeauthoryear{{Kushwaha}, {Gupta}, {Misra}  \&
  {Singh}}{{Kushwaha} et~al.}{2017}]{2017MNRAS.464.2046K}
{Kushwaha} P.,  {Gupta} A.~C.,  {Misra} R.,   {Singh} K.~P.,  2017, \mn@doi
  [\mnras] {10.1093/mnras/stw2440}, \href
  {http://adsabs.harvard.edu/abs/2017MNRAS.464.2046K} {464, 2046}

\bibitem[\protect\citeauthoryear{{Liao}, {Bai}, {Liu}, {Weng}, {Chen}  \&
  {Li}}{{Liao} et~al.}{2014}]{2014ApJ...783...83L}
{Liao} N.~H.,  {Bai} J.~M.,  {Liu} H.~T.,  {Weng} S.~S.,  {Chen} L.,   {Li} F.,
   2014, \mn@doi [\apj] {10.1088/0004-637X/783/2/83}, \href
  {http://adsabs.harvard.edu/abs/2014ApJ...783...83L} {783, 83}

\bibitem[\protect\citeauthoryear{{MacDonald}, {Marscher}, {Jorstad}  \&
  {Joshi}}{{MacDonald} et~al.}{2015}]{2015ApJ...804..111M}
{MacDonald} N.~R.,  {Marscher} A.~P.,  {Jorstad} S.~G.,   {Joshi} M.,  2015,
  \mn@doi [\apj] {10.1088/0004-637X/804/2/111}, \href
  {http://adsabs.harvard.edu/abs/2015ApJ...804..111M} {804, 111}

\bibitem[\protect\citeauthoryear{{Mannheim}}{{Mannheim}}{1993}]{1993A&A...269...67M}
{Mannheim} K.,  1993, \aap, \href
  {http://adsabs.harvard.edu/abs/1993A%26A...269...67M} {269, 67}

\bibitem[\protect\citeauthoryear{{Mao}, {Urry}, {Massaro}, {Paggi},
  {Cauteruccio}  \& {K{\"u}nzel}}{{Mao} et~al.}{2016}]{2016ApJS..224...26M}
{Mao} P.,  {Urry} C.~M.,  {Massaro} F.,  {Paggi} A.,  {Cauteruccio} J.,
  {K{\"u}nzel} S.~R.,  2016, \mn@doi [\apjs] {10.3847/0067-0049/224/2/26},
  \href {http://adsabs.harvard.edu/abs/2016ApJS..224...26M} {224, 26}

\bibitem[\protect\citeauthoryear{{Marscher} \& {Gear}}{{Marscher} \&
  {Gear}}{1985}]{1985ApJ...298..114M}
{Marscher} A.~P.,  {Gear} W.~K.,  1985, \mn@doi [\apj] {10.1086/163592}, \href
  {http://adsabs.harvard.edu/abs/1985ApJ...298..114M} {298, 114}

\bibitem[\protect\citeauthoryear{{Mattox} et~al.,}{{Mattox}
  et~al.}{1996}]{1996ApJ...461..396M}
{Mattox} J.~R.,  et~al., 1996, \mn@doi [\apj] {10.1086/177068}, \href
  {http://adsabs.harvard.edu/abs/1996ApJ...461..396M} {461, 396}

\bibitem[\protect\citeauthoryear{{M{\"u}cke} \& {Protheroe}}{{M{\"u}cke} \&
  {Protheroe}}{2001}]{2001APh....15..121M}
{M{\"u}cke} A.,  {Protheroe} R.~J.,  2001, \mn@doi [Astroparticle Physics]
  {10.1016/S0927-6505(00)00141-9}, \href
  {http://adsabs.harvard.edu/abs/2001APh....15..121M} {15, 121}

\bibitem[\protect\citeauthoryear{{M{\"u}cke}, {Protheroe}, {Engel}, {Rachen}
  \& {Stanev}}{{M{\"u}cke} et~al.}{2003}]{2003APh....18..593M}
{M{\"u}cke} A.,  {Protheroe} R.~J.,  {Engel} R.,  {Rachen} J.~P.,   {Stanev}
  T.,  2003, \mn@doi [Astroparticle Physics] {10.1016/S0927-6505(02)00185-8},
  \href {http://adsabs.harvard.edu/abs/2003APh....18..593M} {18, 593}

\bibitem[\protect\citeauthoryear{{Nolan} et~al.,}{{Nolan}
  et~al.}{2012}]{2012ApJS..199...31N}
{Nolan} P.~L.,  et~al., 2012, \mn@doi [\apjs] {10.1088/0067-0049/199/2/31},
  \href {http://adsabs.harvard.edu/abs/2012ApJS..199...31N} {199, 31}

\bibitem[\protect\citeauthoryear{{Paliya}, {Stalin}  \& {Ravikumar}}{{Paliya}
  et~al.}{2015a}]{2015AJ....149...41P}
{Paliya} V.~S.,  {Stalin} C.~S.,   {Ravikumar} C.~D.,  2015a, \mn@doi [\aj]
  {10.1088/0004-6256/149/2/41}, \href
  {http://adsabs.harvard.edu/abs/2015AJ....149...41P} {149, 41}

\bibitem[\protect\citeauthoryear{{Paliya}, {Sahayanathan}  \&
  {Stalin}}{{Paliya} et~al.}{2015b}]{2015ApJ...803...15P}
{Paliya} V.~S.,  {Sahayanathan} S.,   {Stalin} C.~S.,  2015b, \mn@doi [\apj]
  {10.1088/0004-637X/803/1/15}, \href
  {http://adsabs.harvard.edu/abs/2015ApJ...803...15P} {803, 15}

\bibitem[\protect\citeauthoryear{{Paliya}, {Diltz}, {B{\"o}ttcher}, {Stalin}
  \& {Buckley}}{{Paliya} et~al.}{2016}]{2016ApJ...817...61P}
{Paliya} V.~S.,  {Diltz} C.,  {B{\"o}ttcher} M.,  {Stalin} C.~S.,   {Buckley}
  D.,  2016, \mn@doi [\apj] {10.3847/0004-637X/817/1/61}, \href
  {http://adsabs.harvard.edu/abs/2016ApJ...817...61P} {817, 61}

\bibitem[\protect\citeauthoryear{{Pati{\~n}o-{\'A}lvarez}
  et~al.,}{{Pati{\~n}o-{\'A}lvarez} et~al.}{2018}]{2018MNRAS.479.2037P}
{Pati{\~n}o-{\'A}lvarez} V.~M.,  et~al., 2018, \mn@doi [\mnras]
  {10.1093/mnras/sty1497}, \href
  {http://adsabs.harvard.edu/abs/2018MNRAS.479.2037P} {479, 2037}

\bibitem[\protect\citeauthoryear{{Peterson} et~al.,}{{Peterson}
  et~al.}{2004}]{2004ApJ...613..682P}
{Peterson} B.~M.,  et~al., 2004, \mn@doi [\apj] {10.1086/423269}, \href
  {http://cdsads.u-strasbg.fr/abs/2004ApJ...613..682P} {613, 682}

\bibitem[\protect\citeauthoryear{{Raiteri} et~al.,}{{Raiteri}
  et~al.}{2011}]{2011A&A...534A..87R}
{Raiteri} C.~M.,  et~al., 2011, \mn@doi [\aap] {10.1051/0004-6361/201117026},
  \href {http://adsabs.harvard.edu/abs/2011A%26A...534A..87R} {534, A87}

\bibitem[\protect\citeauthoryear{{Sahayanathan} \& {Godambe}}{{Sahayanathan} \&
  {Godambe}}{2012}]{2012MNRAS.419.1660S}
{Sahayanathan} S.,  {Godambe} S.,  2012, \mn@doi [\mnras]
  {10.1111/j.1365-2966.2011.19829.x}, \href
  {http://adsabs.harvard.edu/abs/2012MNRAS.419.1660S} {419, 1660}

\bibitem[\protect\citeauthoryear{{Sahayanathan}, {Sinha}  \&
  {Misra}}{{Sahayanathan} et~al.}{2018}]{2018RAA....18...35S}
{Sahayanathan} S.,  {Sinha} A.,   {Misra} R.,  2018, \mn@doi [Research in
  Astronomy and Astrophysics] {10.1088/1674-4527/18/3/35}, \href
  {http://adsabs.harvard.edu/abs/2018RAA....18...35S} {18, 035}

\bibitem[\protect\citeauthoryear{{Shah}, {Sahayanathan}, {Mankuzhiyil},
  {Kushwaha}, {Misra}  \& {Iqbal}}{{Shah} et~al.}{2017}]{2017MNRAS.470.3283S}
{Shah} Z.,  {Sahayanathan} S.,  {Mankuzhiyil} N.,  {Kushwaha} P.,  {Misra} R.,
   {Iqbal} N.,  2017, \mn@doi [\mnras] {10.1093/mnras/stx1194}, \href
  {http://adsabs.harvard.edu/abs/2017MNRAS.470.3283S} {470, 3283}

\bibitem[\protect\citeauthoryear{{Sikora}, {Begelman}  \& {Rees}}{{Sikora}
  et~al.}{1994}]{1994ApJ...421..153S}
{Sikora} M.,  {Begelman} M.~C.,   {Rees} M.~J.,  1994, \mn@doi [\apj]
  {10.1086/173633}, \href {http://adsabs.harvard.edu/abs/1994ApJ...421..153S}
  {421, 153}

\bibitem[\protect\citeauthoryear{{Smith}, {Montiel}, {Rightley}, {Turner},
  {Schmidt}  \& {Jannuzi}}{{Smith} et~al.}{2009}]{2009arXiv0912.3621S}
{Smith} P.~S.,  {Montiel} E.,  {Rightley} S.,  {Turner} J.,  {Schmidt} G.~D.,
  {Jannuzi} B.~T.,  2009, preprint, \href
  {http://adsabs.harvard.edu/abs/2009arXiv0912.3621S} {} (\mn@eprint {arXiv}
  {0912.3621})

\bibitem[\protect\citeauthoryear{{Urry} \& {Padovani}}{{Urry} \&
  {Padovani}}{1995}]{1995PASP..107..803U}
{Urry} C.~M.,  {Padovani} P.,  1995, \mn@doi [\pasp] {10.1086/133630}, \href
  {http://cdsads.u-strasbg.fr/abs/1995PASP..107..803U} {107, 803}

\bibitem[\protect\citeauthoryear{{Vercellone} et~al.,}{{Vercellone}
  et~al.}{2009}]{2009ApJ...690.1018V}
{Vercellone} S.,  et~al., 2009, \mn@doi [\apj] {10.1088/0004-637X/690/1/1018},
  \href {http://adsabs.harvard.edu/abs/2009ApJ...690.1018V} {690, 1018}

\bibitem[\protect\citeauthoryear{{Vercellone} et~al.,}{{Vercellone}
  et~al.}{2010}]{2010ApJ...712..405V}
{Vercellone} S.,  et~al., 2010, \mn@doi [\apj] {10.1088/0004-637X/712/1/405},
  \href {http://adsabs.harvard.edu/abs/2010ApJ...712..405V} {712, 405}

\bibitem[\protect\citeauthoryear{{Vercellone} et~al.,}{{Vercellone}
  et~al.}{2011}]{2011ApJ...736L..38V}
{Vercellone} S.,  et~al., 2011, \mn@doi [\apjl] {10.1088/2041-8205/736/2/L38},
  \href {http://adsabs.harvard.edu/abs/2011ApJ...736L..38V} {736, L38}

\bibitem[\protect\citeauthoryear{{Vittorini}, {Tavani}, {Cavaliere}, {Striani}
  \& {Vercellone}}{{Vittorini} et~al.}{2014}]{2014ApJ...793...98V}
{Vittorini} V.,  {Tavani} M.,  {Cavaliere} A.,  {Striani} E.,   {Vercellone}
  S.,  2014, \mn@doi [\apj] {10.1088/0004-637X/793/2/98}, \href
  {http://adsabs.harvard.edu/abs/2014ApJ...793...98V} {793, 98}

\bibitem[\protect\citeauthoryear{{Vovk} \& {Neronov}}{{Vovk} \&
  {Neronov}}{2016}]{2016A&A...586A.150V}
{Vovk} I.,  {Neronov} A.,  2016, \mn@doi [\aap] {10.1051/0004-6361/201526918},
  \href {http://adsabs.harvard.edu/abs/2016A%26A...586A.150V} {586, A150}

\bibitem[\protect\citeauthoryear{{Wagner} \& {Witzel}}{{Wagner} \&
  {Witzel}}{1995}]{1995ARA&A..33..163W}
{Wagner} S.~J.,  {Witzel} A.,  1995, \mn@doi [\araa]
  {10.1146/annurev.aa.33.090195.001115}, \href
  {http://cdsads.u-strasbg.fr/abs/1995ARA%26A..33..163W} {33, 163}

\makeatother
\end{thebibliography}

\bsp	
\label{lastpage}
\end{document}